\documentclass[aps, twocolumn, prb, floatfix, superscriptaddress,longbibliography]{revtex4-2}
\usepackage{amsmath,amssymb,amsthm}
\usepackage{graphicx}
\usepackage[normalem]{ulem}
\usepackage{xcolor}

\newcommand{\Rmnum}[1]{\uppercase\expandafter{\romannumeral #1}}
\usepackage{color}
\definecolor{darkblue}{rgb}{0,0,.65}
\definecolor{darkgreen}{rgb}{0.3,0.6,0.3}
\definecolor{cyan1}{rgb}{0.0, 0.6, 0.6}
\usepackage[%
pdfstartview=FitH,%
breaklinks=true,%
bookmarks=true,%
colorlinks=true,%
anchorcolor=black,%
citecolor=red,
filecolor=black,%
menucolor=black,%
urlcolor=darkblue,%
linkcolor=blue,%
]{hyperref}
\usepackage{cleveref}

\begin{document}

\title{Enhanced Density Fluctuations Near a Disordered Chiral Topological Transition}

\author{Hai-Tao Ding}
\email{htding.9@nus.edu.sg}
\affiliation{Centre for Quantum Technologies, National University of Singapore, 3 Science Drive 2, Singapore 117543}
\affiliation{MajuLab, CNRS-UNS-NUS-NTU International Joint Research Unit, UMI 3654, Singapore}

\author{Sen Mu}
\affiliation{Max Planck Institute for the Physics of Complex Systems, N\"{o}thnitzer Straße 38, 01187 Dresden, Germany}

\author{Leong-Chuan Kwek}
\affiliation{Centre for Quantum Technologies, National University of Singapore, 3 Science Drive 2, Singapore 117543}
\affiliation{MajuLab, CNRS-UNS-NUS-NTU International Joint Research Unit, Singapore UMI 3654, Singapore}
\affiliation{National Institute of Education, Nanyang Technological University, Singapore 637616, Singapore}
\affiliation{Quantum Science and Engineering Centre (QSec), Nanyang Technological University, Singapore}

\author{Gabriel Lemari\'{e}}
\email{gabriel.lemarie@cnrs.fr}
\affiliation{MajuLab, CNRS-UNS-NUS-NTU International Joint Research Unit, Singapore UMI 3654, Singapore}
\affiliation{Centre for Quantum Technologies, National University of Singapore, 3 Science Drive 2, Singapore 117543}
\affiliation{Department of Physics, National University of Singapore, 2 Science Drive 3, Singapore 117551, Singapore}
\affiliation{INPHYNI, Université Côte d’Azur, CNRS, Nice, France}

\author{Jiangbin Gong}
\email{phygj@nus.edu.sg}
\affiliation{Centre for Quantum Technologies, National University of Singapore, 3 Science Drive 2, Singapore 117543}
\affiliation{MajuLab, CNRS-UNS-NUS-NTU International Joint Research Unit, Singapore UMI 3654, Singapore}
\affiliation{Department of Physics, National University of Singapore, 2 Science Drive 3, Singapore 117551, Singapore}

\begin{abstract}
The universal statistics of density fluctuations of localized quantum states may offer unprecedented opportunities to probe and understand quantum transport in connection with dimensionality, coherence, symmetry and disorder. To date, the possible role of topological phase transitions
in the fluctuation statistics is not studied yet. Using a Su-Schrieffer-Heeger chain subject to off-diagonal disorder (so that chiral symmetry is preserved), this work investigates how a disorder driven topological phase transition impacts on the spatial fluctuations of the logarithmic wave-packet density $\ln P(r)$ at distance $r$ from the initial excitation. Away from the transition, in both topological and trivial localized phases, the standard deviation follows the conventional one-dimensional scaling $\sigma[\ln P(r)]\sim r^{\theta}$ with $\theta\simeq 1/2$. Near the transition, however, the fluctuation growth is enhanced: the fitted exponent $\theta$ increases above $1/2$ in a nonmonotonic manner before returning close to $1/2$ at criticality. We interpret this behavior from the energy-resolved density of states and localization length. Near the transition, several energy sectors carry appreciable spectral weight and exhibit competitive decay rates, preventing a single localization scale from dominating the accessible wave-packet tail and thereby enhancing the fluctuations of $\ln P(r)$. Our results establish wave-packet fluctuation statistics as a dynamical indicator of disordered chiral topological transitions and motivate broader studies of fluctuation phenomena in disordered topological quantum systems.

\end{abstract}
\maketitle
\section{Introduction}
\label{sec:introduction}
The interplay of topology and disorder has attracted intense interest in modern condensed matter physics. Disorder can destroy, reshape, or even generate topological phases. A prominent example is the topological Anderson insulator~\cite{li2009topological}, which has recently been realized experimentally in disordered topological systems~\cite{meier2018observation,stutzer2018photonic,liu2020topological,zangeneh2020disorder,zhang2021experimental,liu2021acoustic,lin2022observation,chen2024realization,ren2024realization,li2024mapping,khudaiberdiev2025two,li2025demonstration,mannai2026realization}. These phases are usually characterized through topological invariants, boundary modes, and transport responses~\cite{li2009topological,groth2009theory,jiang2009numerical,guo2010topological,altland2014quantum,wu2016disorder,tang2020topological,roy2021reentrant,zuo2022reentrant,nava2023lindblad,cinnirella2024fate,wang2025emergent,sircar2025disorder,zhang2026reentrant}. However, much less is known about how a disorder-driven topological phase transition may be manifested and probed in the statistical aspects of localized quantum states. This question is particularly natural because Anderson localization is intrinsically statistical: even when the mean profile is exponentially localized, the density or its logarithm can fluctuate strongly from one disorder realization to another.

In disordered quantum systems, universal fluctuation phenomena are often discussed within random-matrix and transfer-matrix frameworks~\cite{beenakker1997random,mello2004quantum,evers2008anderson}. Random-matrix theory provides a statistical description of spectral correlations, scattering matrices, and mesoscopic transport fluctuations after microscopic details are coarse grained~\cite{beenakker1997random,mehta2004random}. In quasi-one-dimensional systems, the Dorokhov-Mello-Pereyra-Kumar equation describes the stochastic evolution of transmission eigenvalues, or the associated Lyapunov exponents, with increasing sample length~\cite{dorokhov1982transmission,mello1988macroscopic,muttalib1999generalized,muttalib2002generalization,mello2004quantum,douglas2014generalized,suslov2018general}. In the localized regime, products of transfer matrices naturally lead to additive fluctuations in logarithmic quantities, including wave-function amplitudes and densities~\cite{mello2004quantum,evers2008anderson}. As a result, the logarithmic density $\ln P(r)$ of localized states exhibits universal scaling behavior. In one or quasi-one dimension, $\ln P(r)$ at large distance $r$ from the localization center scales as $\sigma[\ln P(r)]\sim r^{1/2}$~\cite{mirlin2000statistics,anderson1980new,kramer1993localization,furstenberg1963noncommuting}. In two dimensions, logarithmic density or conductance fluctuations are instead connected to the Kardar-Parisi-Zhang universality class~\cite{kardar1986dynamic} with $\sigma[\ln P(r)]\sim r^{1/3}$~\cite{somoza2007universal,lemarie2019glassy,mu2024kardar,pbrz-3lrl,7d45-yrwq,izem2025kardar}. These results show that fluctuation statistics provide a refined characterization of localized states, beyond the mean exponential decay. Since topology and symmetry impose additional structure on localization, it is natural to ask whether such fluctuation statistics are modified across a disordered topological phase transition. This question is the focus of the present work.

A minimal setting for this question is provided by the one-dimensional Su-Schrieffer-Heeger (SSH) chain with off-diagonal disorder~\cite{su1979solitons}. In Anderson localization with onsite disorder, all eigenstates are exponentially localized in one dimension~\cite{anderson1958absence,lagendijk2009fifty,mott1961theory,abrahams1979scaling,borland1963nature,kramer1993localization}. By contrast, off-diagonal (hopping) disorder preserving chiral symmetry gives the critical zero-energy sector a special role. While eigenstates at non-zero energy remain exponentially localized, the critical zero-energy states can display anomalous behavior, including divergent localization length and sub-exponential typical profiles at criticality~\cite{fleishman1977fluctuations,soukoulis1981off,mondragon2014topological,de2016generalized,dyson1953dynamics,weissmann1975density,bush1975anomalous,theodorou1976extended,eggarter1978singular,ziman1982localization,hilke2025disordered,brouwer2000density,balents1997delocalization}. The coexistence of ordinary exponentially localized finite-energy states and non-trivial (near)-zero-energy states suggests that the spatial fluctuations of a time-evolved wave packet may provide a sensitive probe of the chiral topological transition.

In this work, we study the long-time density profile of a wave packet initially localized in a single unit cell of a disordered SSH chain. We focus on the disorder-induced fluctuations of $\ln P(r)$, an experimentally accessible quantity obtained from the spatial probability profile of the evolved wave packet, as a function of the distance $r$ from the initial excitation. In both the topological and trivial localized phases away from the transition, we find the conventional one-dimensional scaling $\sigma[\ln P(r)]\sim r^{\theta}$ with $\theta\simeq 1/2$. Near the transition, however, the fluctuation growth is enhanced: the numerically fitted exponent rises above $1/2$ in a nonmonotonic manner, before returning close to $1/2$ at the critical point. This behavior shows that the strongest fluctuation enhancement does not occur exactly at criticality, but rather in the near-critical regime. Exactly how the above-mentioned growth exponent of the fluctuation statistics behaves near a topological phase transition will be investigated computationally in this work.

We interpret this nonmonotonic behavior in terms of energy-resolved localization. The two relevant ingredients are the disorder averaged density of states $\langle\rho(E)\rangle$, which determines the density of available states in different energy sectors, and the localization length $\xi(E)$, which determines their decay rates. Away from the transition, the relevant energy sectors have comparable decay rates, so the wave-packet tail is effectively governed by a narrow range of localization scales, leading to the conventional scaling $\sigma[\ln P(r)]\sim r^{1/2}$. Near the transition, the disorder averaged density of states and localization-length spectrum indicate that several energy sectors contribute with appreciable spectral weight and distinct decay rates. As a result, no single localization scale dominates the wave-packet tail over the accessible distances, and this competition enhances the fluctuation growth of $\ln P(r)$. At the critical point, by contrast, the sharp zero-energy sector provides a cleaner dominant contribution, restoring a fluctuation exponent close to the conventional value despite the sub-exponential typical spatial profile. Our results establish wave-packet fluctuation statistics as a dynamical indicator of disordered chiral topological transitions and motivate broader studies of fluctuation phenomena in topological Anderson systems.

The paper is organized as follows. In Sec.~\ref{sec:model}, we define the disordered SSH model, characterize its real-space winding number, and locate the chiral topological transition using both numerical and transfer-matrix diagnostics. In Sec.~\ref{sec:wavepacket}, we present the long-time wave-packet profiles and the fluctuation scaling of $\ln P(r)$ across the topological, topological-Anderson, critical, and trivial regimes. In Sec.~\ref{sec:mechanism}, we develop the energy-resolved interpretation of the enhanced fluctuation exponent in terms of the disorder averaged density of states and localization-length spectrum. We conclude in Sec.~\ref{sec:conclusion} with a summary and outlook.

\section{Disordered SSH model and its phase diagram}
\label{sec:model}
We consider a disordered SSH model with random intracell hopping~\cite{su1979solitons}
\begin{equation}
\label{H_SSH}
H=\sum_{i=1}^Nt_{1,i}a_{i}^{\dagger}b_i + \sum_{i=1}^{N-1}t_{2}a_{i+1}^{\dagger}b_i+\mathrm{H.c.},
\end{equation}
where \(\mathrm{H.c.}\) denotes the Hermitian-conjugate term. $a_{i}^{\dagger}$ $(b_{i}^{\dagger})$ and $a_{i}$ $(b_{i})$ are the creation and annihilation operators for sublattice $A$ $(B)$ in the $i$-th unit cell, respectively. $t_2$ is the intercell hopping strength. $t_{1,i}$ is the intracell hopping strength with $t_{1,i}=t_1+\delta W$, $t_1$ is the mean intracell hopping strength, and $\delta W$ are independent random numbers chosen uniformly
in the range $[-W, W]$. Such hopping disorder breaks the discrete translational invariance of the system, but preserves the chiral symmetry
$\Gamma H \Gamma^{-1} = -H$, where the chiral symmetry operator is defined as $\Gamma=I_N\otimes \sigma_z$, with $I_N$ the $N\times N$ identity matrix and $\sigma_z$ the $2\times 2$ Pauli-$z$ matrix. 

\subsection{Real-space winding number}

\begin{figure*}[t]
  	\centering \includegraphics[width=0.95\textwidth]{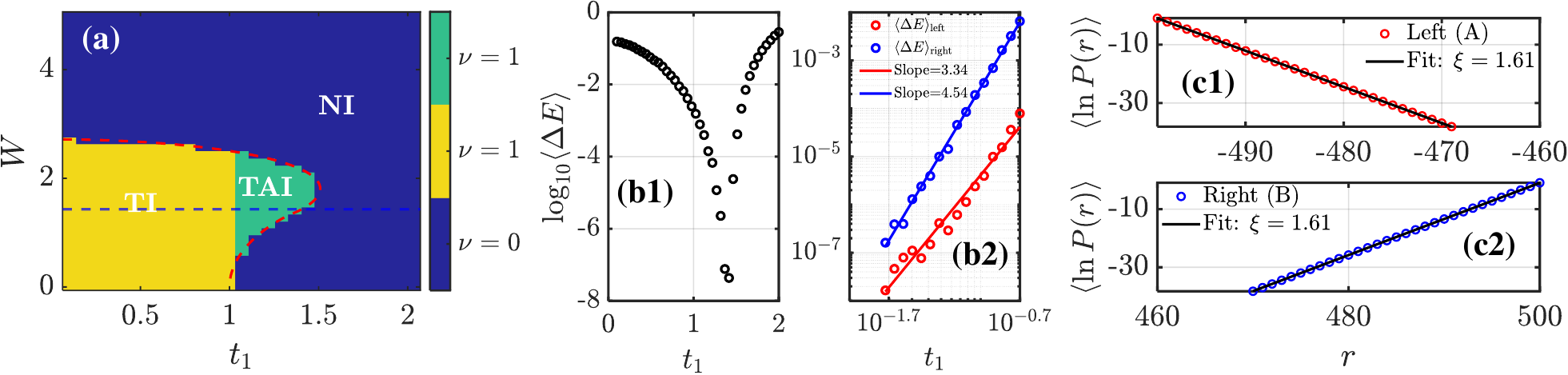}
	\caption{(a) The phase diagram in the $(t_1, W)$ plane. Both TI and TAI regions are characterized by winding number $v=1$, while the NI region has $v=0$. The TAI phase is distinguished from the TI phase in that it is disorder-induced from a regime that is trivial in the clean limit. Analytically derived phase boundary from zero-energy transfer-matrix
recursion is shown with the red dashed line, see ~\cite{asboth2016short,mondragon2014topological} and Appendix~\ref{Ana_Lya} for more details. (b1-b2) Disorder-averaged energy gap $\log_{10}\langle\Delta E\rangle$ as a function of $t_1$, with $W=1.43$. (b1) $t_1\in (0.1,2)$, the gap closing at the transition point $t_1^c=1.40$; (b2)
    Scaling of the disorder averaged energy gap $\langle\Delta E\rangle$ versus $|t_1-t_1^c|$ on a double-logarithmic scale. The left and right branches are shown by red and blue circles, respectively. Solid lines are linear fits in log scale, corresponding to power-law exponents 
3.34 and 4.54.  (c1-c2) Disorder-averaged profiles $\langle\ln P(r)\rangle$ of the left and right topological edge states, with $t_1=0.29$ and $W=1.43$. Here $r$ is the unit-cell index. To display the two edge states in a unified way, the profile of the left-edge state is plotted on the negative-$r$
side, while that of the right-edge state is plotted on the positive-$r$ side. (c1) Profile of the left edge state, localized near the left boundary and residing exclusively on the sublattice A. (c2) Profile of the right edge state, localized near the right boundary and residing exclusively on the sublattice B. Solid lines are exponential fits, yielding the same localization length $\xi=1.61$ for both edge states, which is consistent with the analytical result obtained
from the zero-energy transfer-matrix recursion~\cite{asboth2016short,mondragon2014topological}, see Appendix~\ref{Ana_Lya} for more detail.  Here the intercell hopping strength $t_2=1$, the number of unit cell is $N=10^3$. Results under periodic boundary conditions are shown in (a) and (b), while (c) shows the corresponding result under open boundary conditions.}
	\label{Phase_diagram}
\end{figure*}

The topology of the disordered chain can be characterized by the real space winding number~\cite{mondragon2014topological,tang2022topological}
\begin{equation}
v=\frac{1}{L^{\prime}} \operatorname{Tr}^{\prime}(\Gamma Q[Q, X]) .
\end{equation}
Here, $Q=\sum_{j=1}^N(|j\rangle\langle j|-|\tilde{j}\rangle\langle\tilde{j}|)$, with $|j\rangle$ and  $|\tilde{j}\rangle$ are the conduction band and the valence band respectively, $X$ is the coordinate operator, and $\operatorname{Tr}^{\prime}$ denotes the trace over the central segment of the lattice of length $L^{\prime}=N$, with $N$ is the number of the unit cell.  Unless otherwise specified, we employ the periodic boundary condition (PBC) throughout this paper. The topological phase diagram of the disordered SSH model depicted in Eq.~\eqref{H_SSH} as the function of $W$ and $t_1$ is shown in Fig.~\ref{Phase_diagram}(a).

The lower-left region is the conventional topological insulator (TI) phase with winding number $v=1$. In addition, the phase diagram contains a disorder-induced topological Anderson insulator (TAI) phase~\cite{li2009topological,groth2009theory,jiang2009numerical,guo2010topological,altland2014quantum,tang2020topological,meier2018observation,stutzer2018photonic,zhang2021experimental,liu2020topological,chen2024realization,lin2022observation,ren2024realization}, which also has $v=1$ but emerges from a regime that is trivial in the clean limit. The rest region of Fig.~\ref{Phase_diagram}(a) is the normal insulator (NI) phase with $v=0$.

\subsection{Zero-energy transfer-matrix recursion}

The above phase transition can also be analytically found from the zero-energy transfer-matrix recursion~\cite{kramer1993localization,mondragon2014topological}, which yields
\begin{equation}
\gamma(E=0)  \equiv \lim _{N \rightarrow \infty} \frac{1}{N} \sum_{i=1}^N \ln \left|\frac{t_{2, i}}{t_{1, i}}\right|, 
\label{Lya_expo}
\end{equation}
with $t_{1, i}=t_1+\delta W$ and $t_{2, i}=t_2$. Here $\gamma(E=0)$ is the signed Lyapunov exponent at zero energy~\cite{kramer1993localization,mondragon2014topological}. For the topological phases,  $\gamma(E=0)>0$, and the zero-energy edge states are exponentially localized at the two edges with localization length $\xi(E=0)=1/\gamma(E=0)$; For the trivial insulator phase, $\gamma(E=0)<0$, so the zero-energy solution is non-normalizable and therefore does not represent a physical edge state. The topological phase transition occurs at $\gamma(E=0)=0$, and the localization length diverges~\cite{mondragon2014topological,asboth2016short}. We analytically derive the phase boundary shown in Fig.~\ref{Phase_diagram}(a) with the red dashed line~\cite{mondragon2014topological}; it is consistent with the topological phase boundary numerically determined from the real space winding number, see Appendix~\ref{Ana_Lya} for more details.

\subsection{Disorder-averaged energy gap}

We then calculate the disorder averaged energy gap $\langle\Delta E\rangle$, here
 $\langle \cdots \rangle$ represents the average over disorder realizations.
The results of $\log_{10}\langle\Delta E\rangle$ are shown in Fig.~\ref{Phase_diagram}(b1). The topological Anderson insulator phase is gapped, although the gap remains small. The topological phase transition occurs at the gapless point where the gap closes and reopens, with $t_1^{c}=1.40$ when the disorder strength $W=1.43$ (The blue dashed line in Fig.~\ref{Phase_diagram}(a) is for $W=1.43$). In Fig.~\ref{Phase_diagram} (b2), we present the gap on a log-log scale as a function of the distance to the critical point (CP), $|t_1-t_1^c|$, taking from both sides. The red circles (blue circles) correspond to approaching
$t_1^c$ from $t_1<t_1^c$ ($t_1>t_1^c$): they exhibit power-law scaling $\langle\Delta E\rangle \propto\left|t_1-t_{1}^c\right|^\nu$, with $\nu=3.34$ ($4.54$) when $t_1\rightarrow t_1^c$. We emphasize that the gap closure alone does not necessarily signal a topological transition~\cite{mondragon2014topological}. We therefore identify the topological phase boundary by the vanishing of the zero-energy signed Lyapunov exponent $\gamma(0)$, together with the accompanying change in the real-space winding number.

\subsection{Disorder-averaged logarithmic density of zero-energy edge states}

In Fig.~\ref{Phase_diagram}(c1-c2), we also examine the disorder averaged logarithm of the density distribution $\langle \ln P(r) \rangle$ of two zero-energy edge states with open boundary condition (OBC). $P(r)$ is the total probability density within the $r$-th unit cell, defined as the sum over the density distribution of two sublattices $A$ and $B$: $P(r) = |\psi_A(r)|^2 + |\psi_B(r)|^2$.
The left and right edge states are localized at the corresponding boundaries and reside exclusively on the $A$ and 
$B$ sublattices, respectively. Both zero-energy states under OBC decay with the same localization length, 
$\xi$=1.61, consistent with the analytical result obtained from the zero-energy transfer-matrix recursion in Eq.~\eqref{Lya_expo}~\cite{asboth2016short,mondragon2014topological}, see Appendix~\ref{Ana_Lya} for more details.

\section{Wave-packet localization and fluctuation scaling across the chiral transition}
\label{sec:wavepacket}

For the one-dimensional Anderson model with uncorrelated onsite disorder, all eigenstates are exponentially localized for arbitrarily weak disorder~\cite{anderson1958absence,lagendijk2009fifty,mott1961theory,abrahams1979scaling,borland1963nature,kramer1993localization}. A wave packet launched from a localized initial state therefore remains spatially confined at long times~\cite{kramer1993localization,evers2008anderson,billy2008direct,roati2008anderson}. In the chiral symmetric SSH chain, finite-energy states are also exponentially localized, but the critical zero-energy sector becomes anomalous at the topological transition: its localization length diverges and its typical profile becomes sub-exponential~\cite{soukoulis1981off,mondragon2014topological,de2016generalized}, see Appendix~\ref{App_Eigenstate} for more details. This motivates a direct study of the long-time wave-packet spatial profile and its fluctuations across disorder realizations.

\subsection{Localization of the wave packets}

\begin{figure}[tb]
	\centering
\includegraphics[width=0.48\textwidth]{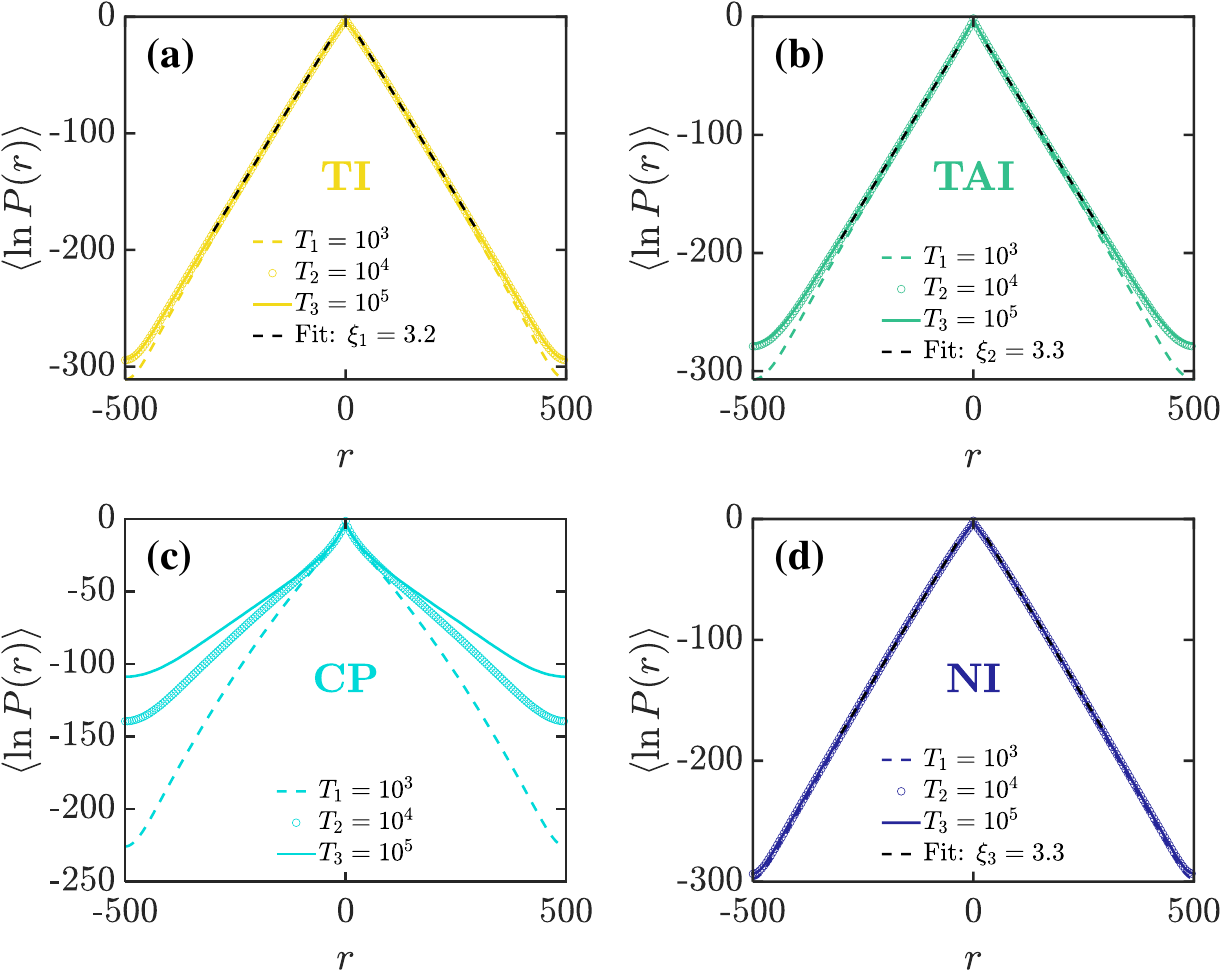}
	\caption{Disorder-averaged logarithm of the wave density $\langle\ln P(r)\rangle$ at three evolution times, $T_1=10^3$, $T_2=10^4$, $T_3=10^5$, for four representative values of $t_1$. (a) $t_1=0.29$ (TI), the profiles at different times nearly collapse and are well fitted by an exponential decay, with the localization length $\xi=3.2$. (b) $t_1=1.12$ (TAI), the profiles also approach a common exponential form at long times, with the localization length $\xi=3.3$.
    (c) The disorder-averaged logarithm of the wave density for $t_1=t_1^c=1.40$ at the critical point (CP). the profile exhibits pronounced time dependence and deviates markedly from a simple exponential form, due to the subexponential decay of the typical profile associated with the $E=0$ eigenstates.
    (d) $t_1=1.90$ (NI), the long-time profile remains exponentially localized, similar to the other insulating phases, with the localization length $\xi=3.3$. 
    Other parameters are $t_2=1$, $N=10^3$,  $W=1.43$, $N_d=10^4$, where $N_d$ denotes the number of disorder realizations.}
	\label{WP_localization}
\end{figure}

We consider an initial state localized at the center of the chain, specifically in the $N/2$-th unit cell, with $N$ the number of unit cells. The state vector is defined as $\psi_{\text{ini}} = (0, \dots, 1/\sqrt{2}, 1/\sqrt{2}, \dots, 0)^\mathcal{T}$. Subsequently, the time evolution is governed by the Hamiltonian in Eq.~\eqref{H_SSH}, under the periodic boundary condition, thereby avoiding any possible artificial effect from the boundary. To investigate the long-time localization behavior of the wave packet, we analyze the disorder-averaged logarithmic density $\langle \ln P(r) \rangle$, where $P(r)=|\psi(r)|^2$ and $r$ denotes the distance from the initial excitation.
We numerically simulate the evolution of the wave packet for four different intracell hopping strength $t_1$ in Fig.~\ref{WP_localization} (a-d): $t_1=0.29$ (deep in TI), $t_1=1.12$ (TAI), $t_1=1.40$ (CP), $t_1=1.90$ (deep in NI).
As shown in Fig.~\ref{WP_localization} (a,b,d), the wave packets exhibit exponential localization $\langle\ln P(r)\rangle\approx-2r/\xi+\text{const}$, with $\xi$ the localization length. Note the absence of $E=0$ eigenstates in these three regimes, so that all eigenstates are expected to be exponentially localized~\cite{soukoulis1981off,mondragon2014topological,de2016generalized}. However, for the case of $t_1=1.40$ shown in Fig.~\ref{WP_localization} (c), the wave packet does not approach a stationary profile even at the longest time considered here ($T_3=10^5$), reflecting some anomalous critical nature of the states, likely connected with the $E=0$ eigenstates at the transition point. At this point, $\gamma(0)=0$ eliminates the linear-in-$r$ contribution to the logarithmic amplitude, leaving fluctuations of order $\sqrt{r}$ and hence the subexponential envelope
$|\psi(r)|_{\mathrm{typ}}\sim\exp(-C\sqrt{|r|})$~\cite{soukoulis1981off,mondragon2014topological,de2016generalized}, where $C>0$ is a nonuniversal constant determined by the strength and statistical properties of the hopping disorder. See also Appendix~\ref{App_Eigenstate} for more details.


\subsection{Fluctuation scaling of the logarithmic wave-packet density as a function of the distance}

We now consider the fluctuation statistics of the logarithmic wave-packet density.  For standard Anderson localization, the fluctuation statistics follows the scaling $\sigma[\ln P(r)]\sim r^{1/2}$ in 1D~\cite{mirlin2000statistics,anderson1980new,kramer1993localization,furstenberg1963noncommuting,mirlin2000statistics}, whereas in 2D it follows the Kardar-Parisi-Zhang scaling $\sigma[\ln P(r)]\sim r^{1/3}$~\cite{kardar1986dynamic,prior2005conductance,prior2009conductance,somoza2015unbinding,somoza2007universal,lemarie2019glassy,mu2024kardar,pbrz-3lrl,izem2025kardar}. Motivated by these results, we examine whether and how the fluctuation scaling is modified in the present chiral-symmetric SSH model across the topological transition. To this end, note that $\langle \ln P(r)\rangle$ describes the typical decay profile, the standard deviation of $\ln P(r)$, 
\begin{equation}
    \sigma[\ln P(r)]=\sqrt{\langle(\ln P(r))^2\rangle-\langle\ln P(r)\rangle^2}.
\end{equation}
quantifies how the distribution broadens across disorder realizations as the distance $r$ increases, and we fit our
numerical results over the spatial range $30\leq r\leq 300$ to
\begin{equation}
\sigma[\ln P(r)]\sim r^\theta .
\end{equation}

\begin{figure}[tb]
	\centering
\includegraphics[width=0.48\textwidth]{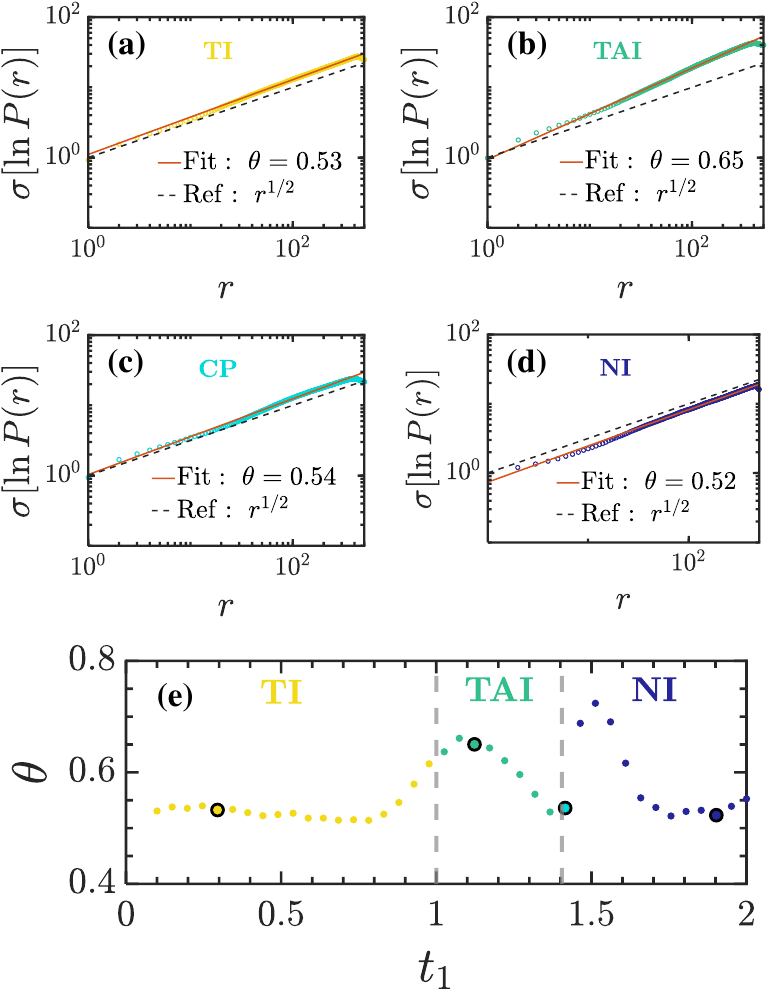}
	\caption{The fluctuation of the logarithm of the wave density $\ln P(r)$ with distance $r$. The blue circles represent the numerical data, the solid red line shows a power-law fit, the black dashed line indicates the  scaling $y = r^{1/2}$ for reference. (a) $t_1=0.29$ (TI), the growth exponent is 0.53. (b) $t_1=1.12$ (TAI), the growth exponent is 0.65. (c) $t_1=1.40$ (CP), the growth exponent is 0.54.
    (d) $t_1=1.90$ (NI), the growth exponent is 0.52. 
    (e) Fitted growth exponent $\theta$ as function of the mean intracell hopping strength $t_1$. The four black open circles in panel (e) mark the four representative values of $t_1$ used in panels (a)–(d).
    The two dashed vertical lines separate the TI, TAI, and NI regimes. Other parameters are $t_2=1$, $N=10^3$, $N_d=10^4$, $W=1.43$.}
	\label{Fluctuation}
\end{figure}

In Fig.~\ref{Fluctuation} (a-d), we present our numerical results for different values of $t_1$ focusing on the fitted growth exponent $\theta$. In the localized TI and NI regimes away from the transition, we find $\theta\simeq 1/2$ once $r$ exceeds the fitted localization length. Thus, despite the chiral symmetry and the distinct topological character of the phases, the long-distance fluctuation scaling agrees with the standard one-dimensional Anderson result. Interestingly, at the critical point $t_1=t_1^{c}$, although the wave packet is not exponentially localized, the fitted fluctuation exponent remains close to $1/2$.

Remarkably, the strongest deviation appears near, but not exactly at, the transition. For $t_1=1.12$ in the TAI regime, the growth exponent is as large as $\theta=0.65$. To further characterize the fluctuation scaling behavior across the topological phase transition, we extract the exponent $\theta$ from the fitted scaling relation $\sigma[\ln P(r)]\sim r^{\theta}$ for different $t_1$. As summarized in Fig.~\ref{Fluctuation} (e), as $t_1$ approaches the critical point ($t_1=t_1^c$), $\theta$ increases above $1/2$ as the transition is approached, but then drops back to a value close to $1/2$ at $t_1=t_1^c$. This nonmonotonic behavior is the central observation of this work. The same nonmonotonic behavior is obtained under open boundary conditions and for larger system sizes up to $N=5000$, as detailed in Appendix~~\ref{App_N=5000}, where an operational characterization of the fluctuation-enhanced regions and a quantitative comparison of their widths for different system sizes are also provided.


\section{Enhanced fluctuation scaling from energy-sector selection}
\label{sec:mechanism}

 
For a one-dimensional localized eigenstate at energy $E$, the logarithmic density at distance $r$ from the localization center can be written as
\begin{equation}
    \ln P_E(r)
    \simeq
    -2r/\xi(E)+\delta h_E(r),
\end{equation}
where $\xi(E)$ is the localization length, and $\delta h_E(r)$ denotes the accumulated random fluctuation of the logarithmic density~\cite{mirlin2000statistics,kramer1993localization,furstenberg1963noncommuting}. For a fixed energy sector, this fluctuation obeys the conventional one-dimensional scaling
    $\sigma[\delta h_E(r)]\sim r^{1/2}$~\cite{mirlin2000statistics,anderson1980new,kramer1993localization,furstenberg1963noncommuting}.
For a time-evolved wave packet, the long-distance tail is determined by the energy sectors selected by their spectral weight and localization length $\xi(E)$.
If several sectors with distinct localization lengths contribute simultaneously, the tail is no longer governed by a single decay channel, and the fluctuation of $\ln P(r)$ is enhanced.
\begin{figure}[tb]
	\centering
\includegraphics[width=0.48\textwidth]{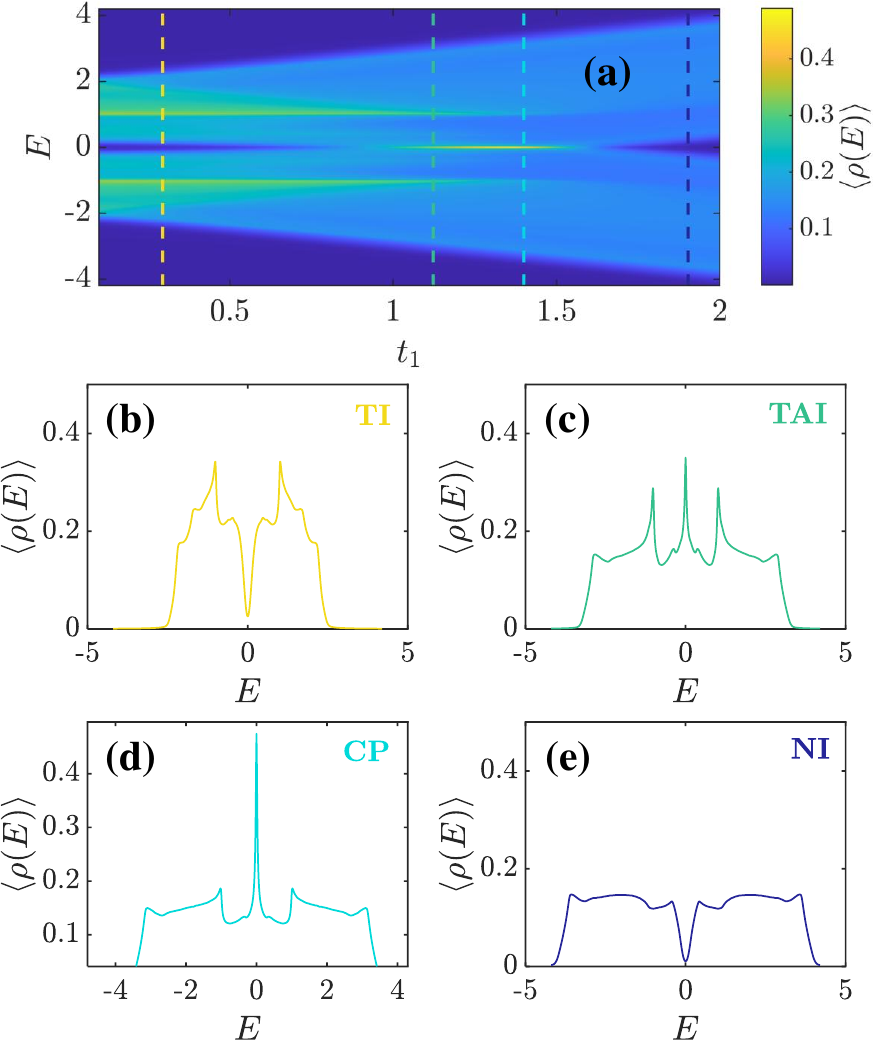}
	\caption{Disorder-averaged density of states $\langle\rho(E)\rangle$ across the chiral topological transition.
(a) $\langle\rho(E)\rangle$ as a function of the mean intracell hopping strength $t_1$ and energy $E$. The four vertical dashed lines mark four representative values of $t_1$ used in panels (b)--(e).
(b)--(e) Energy dependence of $\langle\rho(E)\rangle$ for $t_1=0.29$ (TI), $t_1=1.12$ (TAI), $t_1=1.40$ (CP), and $t_1=1.90$ (NI), respectively. At $t_1 = 1.12$, the system is still gapped, and the central peak corresponds to near-zero-energy states rather than states exactly at $E = 0$. At the critical point $t_1 = 1.40$, the gap closes and the sharp peak occurs at $E = 0$. In the near-critical TAI regime, spectral weight appears both near zero energy and finite-energy sectors, indicating the coexistence of several relevant energy channels.
At the critical point, the density of states is dominated by a sharp zero-energy contribution.
 Other parameters are $t_2 = 1$, $N=10^3$, $N_d = 10^4$, and $W = 1.43$.}
	\label{DOS}
\end{figure}

\begin{figure}[tb]
	\centering
\includegraphics[width=0.48\textwidth]{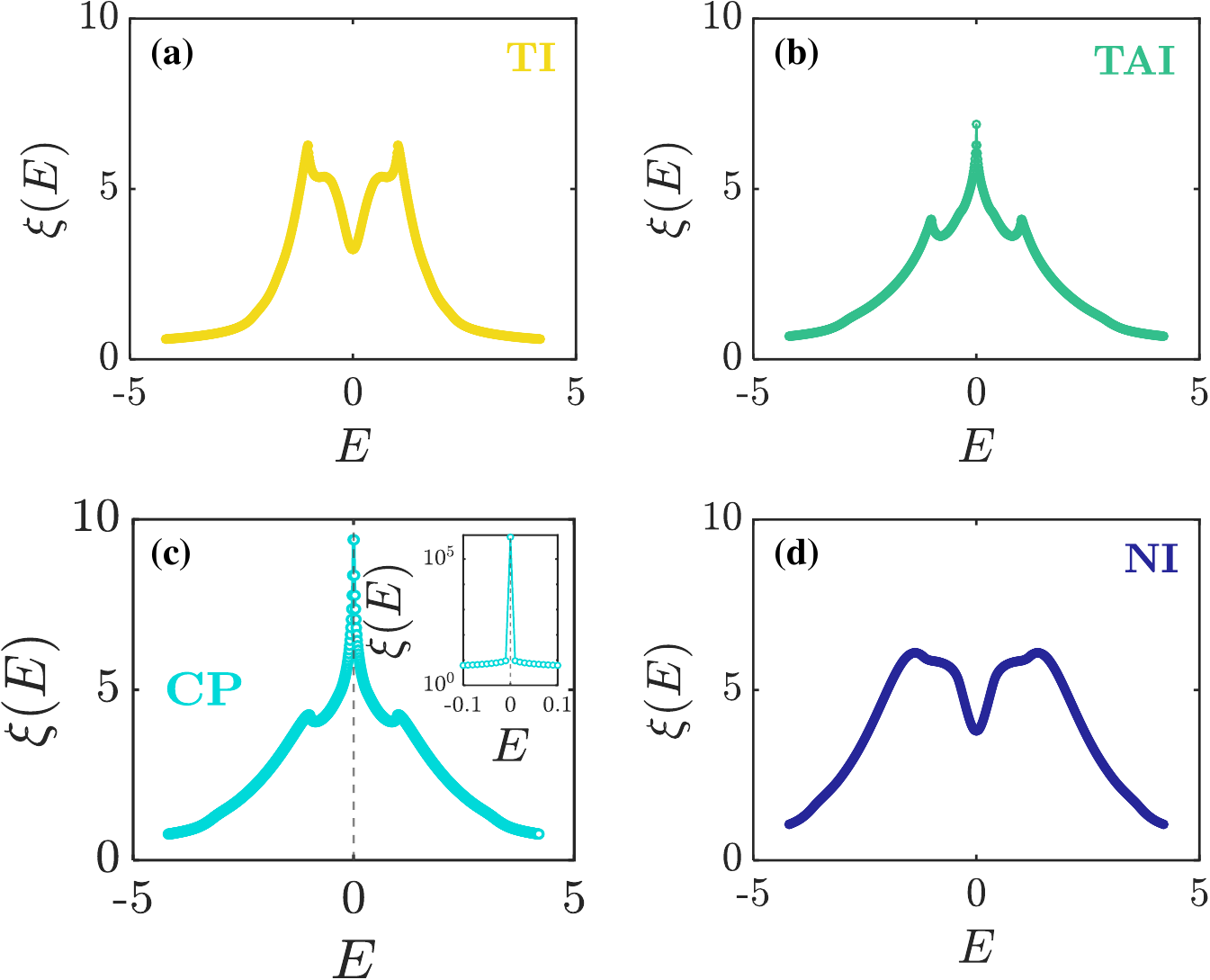}
	\caption{The energy-resolved localization length $\xi(E)$ for four representative values of $t_1$. (a) $t_1=0.29$ (TI). (b) $t_1=1.12$ (TAI). (c) $t_1=1.40$ (CP). (d) $t_1=1.90$ (NI). The localization length is obtained from the inverse magnitude of the disorder-averaged Lyapunov exponent.
   In the TI and NI regimes, $\xi(E)$ remains finite over the relevant energy range, indicating that the long-distance wave-packet tail is governed by a stable effective decay scale. In the near-critical TAI regime, several energy sectors have comparable but distinct localization lengths, so the tail is not selected by a single decay channel; this competition accounts for the enhanced fluctuation of $\ln P(r)$.  
   At the critical point, $\xi(E)$ diverges at $E=0$, reflecting the vanishing zero-energy Lyapunov exponent and showing that the zero-energy sector controls the long-distance dynamics. In panel (c), the main plot uses a linear scale to display the finite-energy structure, whereas the inset shows the same data near $E=0$ on a logarithmic scale, highlighting the divergent zero-energy localization length. Other parameters are $t_2 = 1$, $N_d = 10^4$, and $W = 1.43$.}
	\label{Loc_length}
\end{figure}

This motivates an energy-resolved analysis of the competing sectors. We therefore compute the disorder-averaged density of states $\langle\rho(E)\rangle=\langle \frac{1}{2N}\sum_n\delta(E-E_n)\rangle$ and the energy-resolved localization length $\xi(E)$, shown in Fig.~\ref{DOS} and Fig.~\ref{Loc_length}, respectively.
In the deep topological localized regime, represented by $t_1=0.29$, the disorder averaged density of states [Fig.~\ref{DOS}(b)] is concentrated in finite-energy regions ($E\approx\pm1$), while the localization-length spectrum [Fig.~\ref{Loc_length}(a)] selects a well-defined finite-energy localization scale ($\xi(E=1)=\xi(E=-1)$). Although many eigenstates contribute to the wave packet, its long-distance tail is effectively governed by a single dominant decay channel. The fluctuation of $\ln P(r)$ therefore follows the ordinary one-dimensional scaling $\sigma[\ln P(r)]\sim r^{1/2}$.
 
The situation is different in the near-critical topological Anderson regime, represented by $t_1=1.12$ [Fig.~\ref{DOS}(c) and Fig.~\ref{Loc_length}(b)]. In this regime, the system remains gapped, but spectral weight accumulates near zero energy while finite-energy sectors remain appreciable. The corresponding localization lengths also vary significantly among these energy windows. As a result, the wave-packet tail is not selected by a single localization scale. Instead, several energy sectors with distinct localization lengths contribute simultaneously, which enhances the growth of $\sigma[\ln P(r)]$ and gives rise to an apparent exponent larger than $1/2$.

The critical point, $t_1=t_1^c=1.40$, should be distinguished from the near-critical competition regime discussed above. At criticality, both $\langle\rho(E)\rangle$ and $\xi(E)$ exhibit a sharply resolved zero-energy sector, as shown in Fig.~\ref{DOS}(d) and Fig.~\ref{Loc_length}(c). The gap closes and the localization length $\xi(E=0)$ diverges, so the long-time wave-packet tail is dominated solely by the critical zero-energy sector. At this critical point, we observe $\sigma[\ln P(r)]\sim r^{1/2}$. This scaling may be inferred from the zero-energy-state analysis of one-dimensional random-hopping chains~\cite{mondragon2014topological,asboth2016short,krishna2020beyond}.

Deep in the trivial localized phase  ($t_1=1.90$), the zero-energy singularity is absent, and all relevant localization lengths remain finite. The disorder averaged density of state [Fig.~\ref{DOS}(e)] varies smoothly over relevant energy window, while $\xi(E)$ [Fig.~\ref{Loc_length}(d)] shows only moderate energy dependence. As a result, the wave-packet tail is not affected by a sharp accumulation of zero-energy states or by a broad competition among different localization scales. The decay is therefore effectively governed by a relatively narrow range of localization lengths, and the usual one-dimensional localization fluctuation scaling $\sigma[\ln P(r)]\sim r^{1/2}$, is recovered.

The resulting picture is therefore not that the fluctuation exponent is controlled simply by the magnitude of the localization length or by the accumulation of near-zero energy states alone. Instead, it is determined by whether the density of states and the energy-resolved localization length favor a narrow range of effective decay channels, or whether several channels with comparable spectral weight and distinct decay rates compete. Away from the transition, the wave-packet tail is effectively controlled by a narrow range of localization scales. At the critical point, the sharp zero-energy sector provides the dominant long-distance contribution. The strongest enhancement appears in the near-critical gapped regime, where near-zero and finite-energy sectors coexist with appreciable spectral weight and distinct localization lengths. Their simultaneous contribution prevents the accessible tail from being governed by a narrow set of effective decay channels, thereby enhancing the fluctuation growth of $\ln P(r)$.


\section{Conclusion and discussion}
\label{sec:conclusion}

We have studied the spatial fluctuation statistics of a long-time wave packet in a disordered SSH chain with chiral symmetry. The central observable is the logarithmic wave packet density $\ln P(r)$, whose standard deviation characterizes the sample-to-sample fluctuations of localized wave packet tails. In the topological and trivial localized phases away from the transition, the fitted effective exponent remains close to the conventional one-dimensional value of $1/2$. 
This indicates that topology alone does not necessarily modify the scaling of logarithmic density fluctuations when the wave-packet tail is governed by a relatively narrow range of localization scales.
However, the the fitted exponent is enhanced in the vicinity of the chiral topological transition, but returns close to $1/2$ at the critical point. 
 
The enhancement originates from the energy-resolved structure of the localized states. In the near-critical regime, localized components with different decay lengths contribute simultaneously to the wave-packet tail, giving rise to a multi-scale decay structure that amplifies the fluctuations of $\ln P(r)$. At criticality, by contrast, the zero-energy sector dominates the long-distance dynamics. As a result, the fitted fluctuation exponent returns close to $1/2$, even though the wave-packet profile itself remains anomalous.
 
Several directions naturally follow from these results. 
One direction is to extend the analysis to disordered systems in different symmetry classes, with the aim of identifying which aspects of the enhanced fluctuation scaling are specific to chiral symmetry and which reflect more general consequences of symmetry-controlled criticality. Another important direction is to study higher-dimensional chiral and topological Anderson systems. 
Two-dimensional chiral random-hopping systems are particularly 
promising, since their near-zero-energy sector can exhibit a singular 
density of states and a diverging localization length, whereas 
finite-energy states remain localized 
\cite{gade1993anderson,eilmes1998two,motrunich2002particle,schweitzer2012scaling}. 
The resulting competition between low and finite energy contributions 
with widely separated decay lengths may introduce additional 
sample-to-sample fluctuations beyond the intrinsic 
Kardar--Parisi--Zhang contribution, potentially driving $\theta$ away 
from $1/3$.

\begin{acknowledgments}
The authors thank Christian Miniatura, Maxime Richard, Hui Khoon Ng, and Zhixing Zou for helpful discussions. We acknowledge the use of computational resources at the Singapore National Super Computing Centre (NSCC) ASPIRE-2A cluster, at Calcul en Midi-Pyrénées (CALMIP) for our simulations. This work is supported by the Singapore Ministry of Education Academic Research Funds Tier II (MOE-T2EP50223-0009 and MOE-T2EP50222-0005), by France 2030 under the French National Research Agency Award QUTISYM No. ANR-23-PETQ-0002 and by the ANR Research Grants ManyBodyNet No. ANR-24-CE30-5851. J.G. also acknowledges support by the National Research Foundation, Singapore, through the National Quantum Office, hosted in A*STAR, under its Centre for Quantum Technologies Funding Initiative (S24Q2d0009).
\end{acknowledgments}

\appendix

\section{Analytical Derivation of $\gamma(0)$ from the $E=0$ transfer matrix recursion}
\label{Ana_Lya}

We provide a rigorous derivation of $\gamma(0)$ for the 1D SSH model with off-diagonal disorder~\cite{mondragon2014topological,asboth2016short}. We first introduce the general transfer matrix formalism and then demonstrate how the zero-energy condition simplifies the problem, allowing for an exact analytical solution.
We consider the one-dimensional SSH model with uncorrelated   intracell hopping~\cite{su1979solitons,meier2018observation}, defined as:
\begin{equation}
    H = \sum_{n} \left( t_{1,n} |n,A\rangle\langle n,B| + t_{2,n} |n,B\rangle\langle n+1,A| + \text{H.c.} \right),
\end{equation}
where $|n,A\rangle$ and $|n,B\rangle$ denote the states on the two sublattices ($A$ and $B$) in the $n$-th unit cell. $t_{1,n}$ represents the intra-cell hopping, and $t_{2,n}$ represents the inter-cell hopping. The Schrödinger equation $H|\psi\rangle = E|\psi\rangle$ for a state $|\psi\rangle = \sum_n (a_n |n,A\rangle + b_n |n,B\rangle)$ yields the following coupled equations for the amplitudes $a_n$ and $b_n$:
\begin{align}
    E a_n &= t_{1,n} b_n + t_{2,n-1} b_{n-1}, \label{eq:Sch_A} \\
    E b_n &= t_{1,n} a_n + t_{2,n} a_{n+1}. \label{eq:Sch_B}
\end{align}
At zero energy ($E=0$), a significant simplification occurs due to the chiral (sublattice) symmetry of the bipartite lattice.
Setting $E=0$ in Eqs.~(\ref{eq:Sch_A}) and (\ref{eq:Sch_B}), the equations for the two sublattices decouple completely:
\begin{align}
    0 &= t_{1,n} b_n + t_{2,n-1} b_{n-1} \implies b_n = -\frac{t_{2,n-1}}{t_{1,n}} b_{n-1}, \label{eq:recursive_b} \\
    0 &= t_{1,n} a_n + t_{2,n} a_{n+1} \implies a_{n+1} = -\frac{t_{1,n}}{t_{2,n}} a_n.
\end{align}
This decoupling implies that the zero-energy eigenstates are confined to either the A-sublattice or the B-sublattice. 
By iterating these two relations from the first unit cell to the $N$-th unit cell, the amplitudes $a_N$ and $b_N$ are expressed as a product of hopping ratios:
\begin{equation}
\begin{aligned}
    a_N & =  \prod_{i=1}^{N-1} \left(-\frac{t_{1,i}}{t_{2,i}}  \right) a_1.   \\
    b_N & =  \prod_{i=2}^{N} \left(-\frac{t_{2,i-1}}{t_{1,i}}  \right) b_1.   
\end{aligned}
\end{equation}
In the zero-energy transfer-matrix recursion, the amplitude ratio defines a signed drift parameter, or equivalently a signed Lyapunov exponent, $\gamma(0)$ as
\begin{equation}
    \gamma(0) \equiv -\lim_{N \to \infty} \frac{1}{N} \ln \left| \frac{a_N}{a_1} \right|.
\end{equation}
Equivalently, it can also be written as
\begin{equation}
    \gamma(0) \equiv \lim_{N \to \infty} \frac{1}{N} \ln \left| \frac{b_N}{b_1} \right|.
\end{equation}
The two expressions are equivalent in the thermodynamic limit, since they describe the same zero-energy bulk drift from opposite recursion directions.
The maximal Lyapunov exponent at $E=0$ is $\lambda(0)=|\gamma(0)|$~\cite{kramer1993localization,mondragon2014topological}.
The associated characteristic length is $\xi(0)=1/\lambda(0)$. When there is a normalizable zero-energy edge mode, $\xi(0)$ coincides with their penetration depth. Substituting the product form into the definition, the logarithm converts the product into a sum:
\begin{equation}
\begin{aligned}
    \gamma(0)&= \lim_{N \to \infty} \frac{1}{N} \sum_{i=1}^{N-1} \ln \left| \frac{t_{2,i}}{t_{1,i+1}} \right|  \\
    &= \lim_{N \to \infty} \frac{1}{N} \sum_{i=1}^{N} \ln \left| \frac{t_{2,i}}{t_{1,i}} \right|.
    \label{eq:sum_log}
\end{aligned}
\end{equation}
We consider the specific case where the inter-cell hopping is uniform ($t_{2,i} = t_2 = 1$) and the intra-cell hopping possesses box-distributed disorder: $t_{1,i} = t_1 + \delta W_i$, where $\delta W_i$ is a random variable uniformly distributed in the interval $[-W, W]$ with probability density function $P(x) = \frac{1}{2W}$.

According to the Birkhoff ergodic theorem, in the thermodynamic limit ($N \to \infty$), the spatial average over the chain is equivalent to the ensemble average over the disorder distribution. Thus, Eq.~(\ref{eq:sum_log}) becomes:
\begin{equation}
    \gamma(0) =- \langle \ln |t_1 + \delta W| \rangle = -\int_{-W}^{W} P(x) \ln |t_1 + x| \, dx.
\end{equation}
The integral can be evaluated analytically, yielding
\begin{equation}
    \gamma(0)= \frac{-(t_1+W) \ln|t_1+W| + (t_1-W) \ln|t_1-W|}{2W} + 1.
\end{equation}
This analytical result provides the $\gamma(0)$ as a function of the disorder strength $W$ and the hopping parameter $t_1$~\cite{mondragon2014topological}.

For the topological insulator phase ($\langle \ln |t_{2,i}| \rangle > \langle \ln |t_{1,i}| \rangle$), we have $\gamma(0) > 0$. This positive exponent implies that the $E=0$ eigenstate decays as $|\psi| \sim e^{-\gamma(0)\cdot x}$. Thus, the zero energy states exponentially localize at the two edges; these are the topologically protected edge states, with the localization length $\xi=1/|\gamma(0)|$. For the trivial insulator phase ($\langle \ln |t_{2,i}| \rangle < \langle \ln |t_{1,i}| \rangle$), we have $\gamma(0) < 0$. This means that any solution to the $E=0$ eigenstates behaves as $|\psi| \sim e^{-\gamma(0)\cdot x} = e^{|\gamma|x}$, which grows exponentially in space. Such an exponentially diverging wave function is non-normalizable and therefore does not correspond to a physically acceptable eigenstate. Consequently, no zero energy states exist in the trivial insulator phase.
For $\gamma(0)=0$, the localization length for zero energy is divergent, indicating that the (formerly) zero-energy edge state delocalizes and merges with the bulk states, which is the topological phase transition point.
So the transition occurs for $\gamma(0)=0$.
This analytical result yields the critical phase boundary, shown as the red dashed line in Fig.~\ref{Phase_diagram}(a) of the main text, which is fully consistent with the topological phase transition calculated from the real space winding number.  For the disorder strength $W=1.43$ considered in the main text, the transition occurs at $t_1^c=1.40$.

For the parameters considered in Fig.~\ref{Phase_diagram}(c1,c2) in the main text: $t_1=0.29$, $W=1.43$, the zero-energy edge states in open boundary condition behave as $\langle\ln|\psi|^2\rangle\sim -2r/\xi$, with the localization length $\xi=1/\gamma(0)=1.61$, which is the same as the numerical results in the main text.

\section{Subexponential eigenstate profiles at the chiral topological transition}
\label{App_Eigenstate}

\subsection{Analytical derivation}

We now derive the subexponential typical profile of the
zero-energy state at the critical point, following the
analysis of Ref.~\cite{krishna2020beyond}. At $E=0$, the equations on the two
sublattices decouple. For a zero-energy solution supported
on sublattice $A$, the recursion relation is
\begin{equation}
a_{n+1}
=
-\frac{t_{1,n}}{t_2}a_n .
\end{equation}
Iterating this relation over $r$ unit cells gives
\begin{equation}
a_{n+r}
=
(-1)^r a_n
\prod_{i=n}^{n+r-1}
\frac{t_{1,i}}{t_2}.
\end{equation}
Taking the logarithm of the absolute value yields
\begin{equation}
\ln\left|
\frac{a_{n+r}}{a_n}
\right|
=
\sum_{i=n}^{n+r-1}
\ln\left|
\frac{t_{1,i}}{t_2}
\right|.
\end{equation}
To connect this expression with the signed Lyapunov
exponent introduced in Appendix~\ref{Ana_Lya}, we define
\begin{equation}
u_i
\equiv
\ln\left|
\frac{t_2}{t_{1,i}}
\right|,
\qquad
\gamma(0)=\langle u_i\rangle ,
\end{equation}
so that
\begin{equation}
\ln\left|
\frac{a_{n+r}}{a_n}
\right|
=
-\sum_{i=n}^{n+r-1}u_i .
\end{equation}
Introducing
$\sigma_u^2=
\left\langle
\left[u_i-\gamma(0)\right]^2
\right\rangle$,
the independence of the intracell hopping disorder gives
\begin{equation}
\begin{aligned}
&\left\langle
\ln\left|
\frac{a_{n+r}}{a_n}
\right|
\right\rangle
=
-r\gamma(0),
\\
&\left\langle
\left[
\ln\left|
\frac{a_{n+r}}{a_n}
\right|
-\left\langle
\ln\left|
\frac{a_{n+r}}{a_n}
\right|
\right\rangle
\right]^2
\right\rangle
=
r\sigma_u^2.
\end{aligned}
\label{var}
\end{equation}
For the box disorder considered here, $\sigma_u^2<\infty$.
At the topological transition, $\gamma(0)=0$.
The mean linear drift of the logarithmic amplitude therefore vanishes, and Eq.~\eqref{var} gives
\begin{equation}
\left|
\ln\left|
\frac{a_{n+r}}{a_n}
\right|
\right|_{\mathrm{typ}}\sim\sigma_u\sqrt{r},
\end{equation}
Taking the global intensity maximum as $r=0$, we have
\begin{equation}
|\psi(r)|_{\mathrm{typ}}
\sim
\exp\left(-C\sqrt{|r|}\right), \quad C>0.
\end{equation}
Since the zero-energy state resides on only one sublattice
and
$P(r)=|\psi_A(r)|^2+|\psi_B(r)|^2$,
the corresponding typical density profile is
\begin{equation}
P_{\mathrm{typ}}(r)
\equiv
\exp\left[\langle\ln P(r)\rangle\right]
\sim
\exp\left(-2C\sqrt{|r|}\right).
\end{equation}
Thus, the critical zero-energy state exhibits
subexponential decay. A zero-energy state supported on sublattice $B$ obeys the analogous recursion and therefore exhibits the same statistical scaling and typical decay envelope.

Away from the transition, \(\gamma(0)\neq0\), and the
linear term \(r\gamma(0)\) dominates over the
\(\sqrt{r}\) fluctuations. The usual exponential envelope
is therefore recovered,
\begin{equation}
|\psi(r)|_{\mathrm{typ}}
\sim
\exp\left[-|\gamma(0)|\,|r|\right].
\end{equation}
Hence, the disappearance of the linear logarithmic drift
at $\gamma(0)=0$ is precisely the origin of the
subexponential zero-energy profile.

\subsection{Numerical verification}
In this part, we discuss the properties of the eigenstates with eigenenergies $E\neq 0$ and $E=0$. For the one-dimensional chiral class at criticality, the eigenstate with eigenenergy $E\neq 0$ show exponential decay $|\psi|\sim e^{-r/\xi}$ while for eigenstates with eigenenergy $E=0$, their typical profile exhibits subexponential decay~\cite{soukoulis1981off,theodorou1976extended,ziman1982localization,hilke2025disordered,balents1997delocalization}.
\begin{figure}[tb]
	\centering
\includegraphics[width=0.48\textwidth]{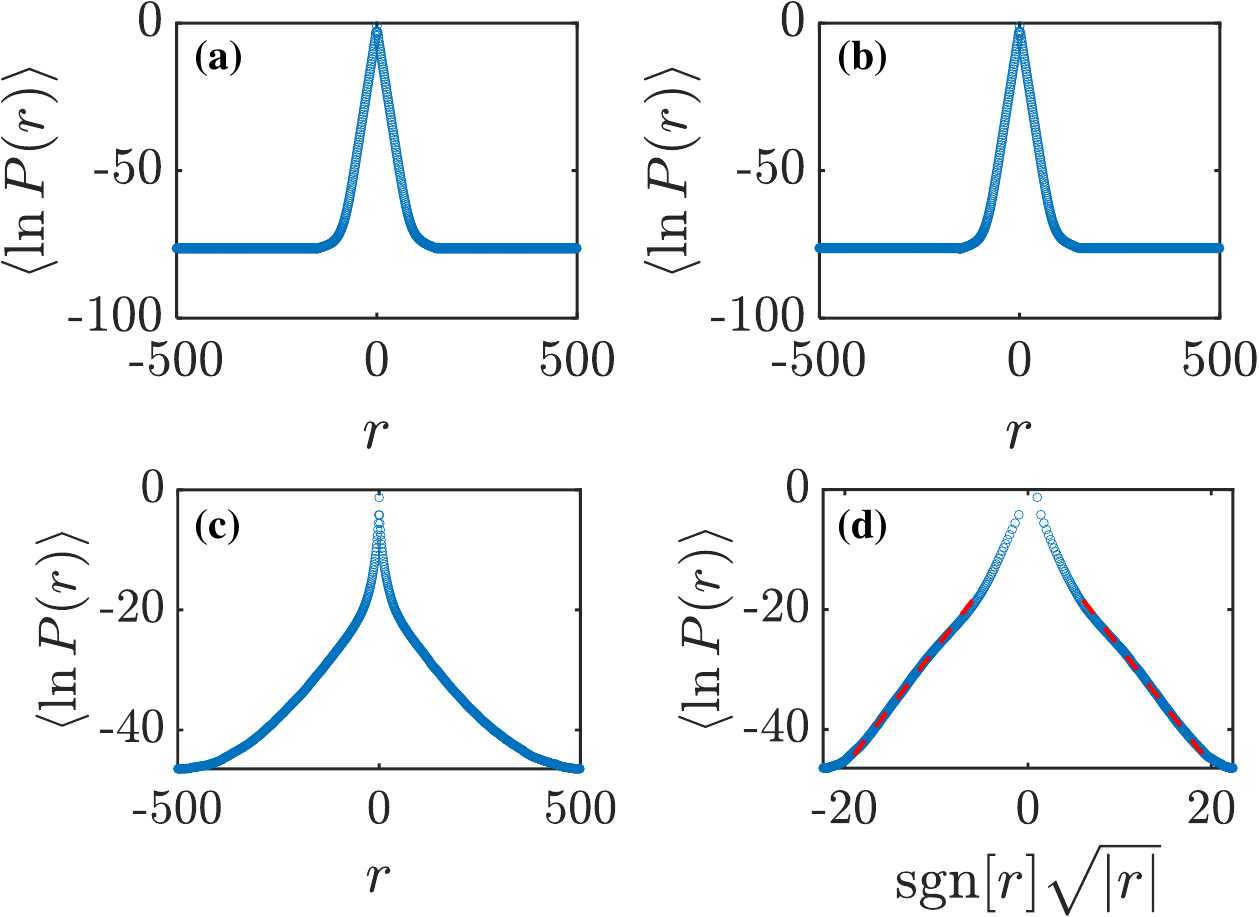}
	\caption{The logarithm of the eigenstates $\langle\ln P(r)\rangle$ for eigenenergies windows, (a) for $E\in(0.40, 0.41)$, it exhibits exponential decay. The plateaus are numerical artifacts arising from the machine precision limit of the diagonalization algorithm. (b) for $E\in(-0.41, -0.40)$, it exhibits exponential decay. The plateaus are numerical artifacts arising from the machine precision limit of the diagonalization algorithm. (c) The logarithm of the eigenstates $\langle\ln P(r)\rangle$ for eigenenergies windows $E\in(-10^{-8}, 10^{-8})$ versus the linear distance $r$ and (d) $\text{sgn}[r]\sqrt{|r|}$.  It exhibits a significant departure from the standard exponential decay. Instead, the decay profile is better characterized by subexponential decay.  Other parameters are $t_1=1.40$, $t_2=1$, $N=1000$, $N_d=10^4$, $W=1.43$. Here we consider the periodic boundary condition.}
	\label{Eigenstate_Loc}
\end{figure}
The localization center for every eigenstates are different, we move all the localization centers of the eigenstates to the center unit cell then calculate disorder averaged logarithm of the eigenstates $\langle\ln|\psi(r)|^2\rangle$.
We choose three different eigenenergy windows $(0.40, 0.41)$, $(-0.41, -0.40)$, and $(-10^{-8}, 10^{-8})$ for $t_1=t_1^c=1.40$. The results of $\langle\ln|\psi|^2\rangle$ are shown 
 in Fig~\ref{Eigenstate_Loc}, the eigenstates with $E \neq 0$ ($E\in(0.40, 0.41)$ and $E\in(-0.41,-0.40)$ shown in Fig~\ref{Eigenstate_Loc}(a,b)) exhibit exponential decay with finite localization length. For eigenstates with eigenenergy $E\rightarrow0$ $(E\in(-10^{-8},10^{-8}))$, the localization length is divergent~\cite{soukoulis1981off,mondragon2014topological}. The  logarithm of density $\langle\ln P(r)\rangle$ as function of $r$ and $\text{sgn}[r]\sqrt{|r|}$ are shown in~\ref{Eigenstate_Loc}(c) and (d), respectively.  It obviously deviated from the exponential decay; it exhibits subexponential decay. The plateaus in Fig~\ref{Eigenstate_Loc}(a,b)) are numerical artifacts arising from the machine precision limit of the diagonalization algorithm. 


\newpage
\section{More on wavepacket dynamics} 
\label{App_N=5000}
\subsection{Open boundary conditions ($N=1000$)}
\begin{figure}[!htbp]
	\centering
\includegraphics[width=0.48\textwidth]{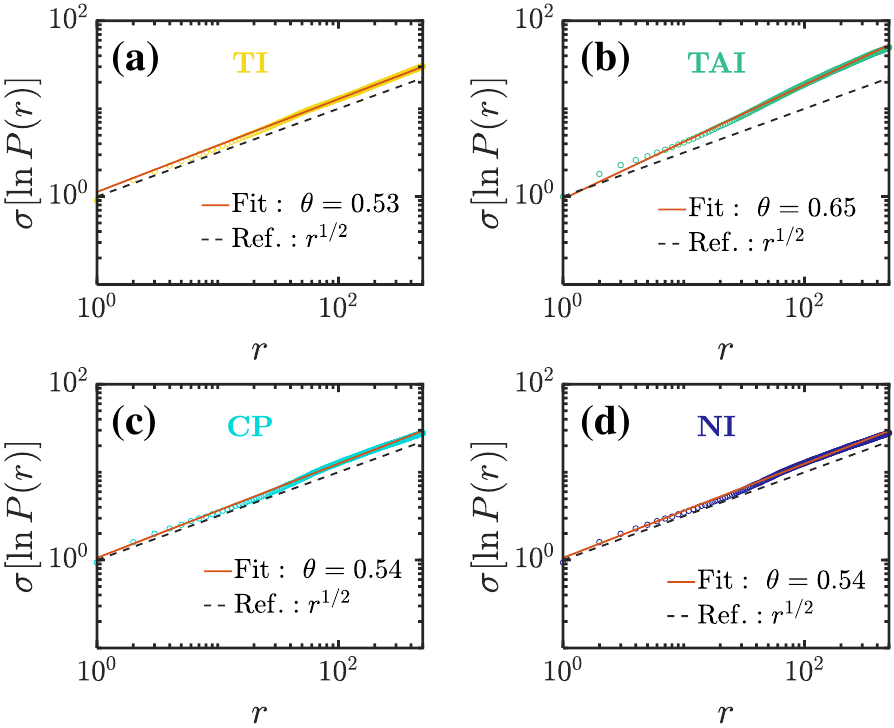}
	\caption{The fluctuation of the logarithm of the wave density $\ln P(r)$ with distance $r$. The blue circles represent the numerical data, the solid red line shows a power-law fit, the black dashed line indicates the  scaling $y = r^{1/2}$ for reference.  (a) $t_1=0.29$ (TI), the growth exponent is 0.53. (b) $t_1=1.12$ (TAI), the growth exponent is 0.65. (c) $t_1=1.40$ (CP) at the transition point, the growth exponent is 0.54. (d) $t_1=1.90$ (NI), the growth exponent is 0.54. Other parameters are, $t_2=1$, $N=1000$, $N_d=10^4$, $W=1.43$. The evolution time is $T=10^4$. Here we consider the open boundary condition.}
	\label{OBC_Fluctuation}
\end{figure}
The fluctuation scaling is essentially unchanged under OBC, as shown in Fig.~\ref{OBC_Fluctuation}. Deep in the TI and NI phases, as
well as at the critical point, $\sigma[\ln P(r)]$ follows the conventional
$r^{1/2}$ scaling, with fitted exponents $\theta=0.53$, $0.54$, and $0.54$,
respectively. By contrast, in the TAI regime close to the transition, the
exponent is enhanced to $\theta=0.65$. These results are consistent with
those obtained under PBC, demonstrating that the enhanced fluctuation
scaling is insensitive to the choice of boundary conditions.

\subsection{Periodic boundary conditions ($N=5000$)}
\begin{figure}[tb]
	\centering
\includegraphics[width=0.48\textwidth]{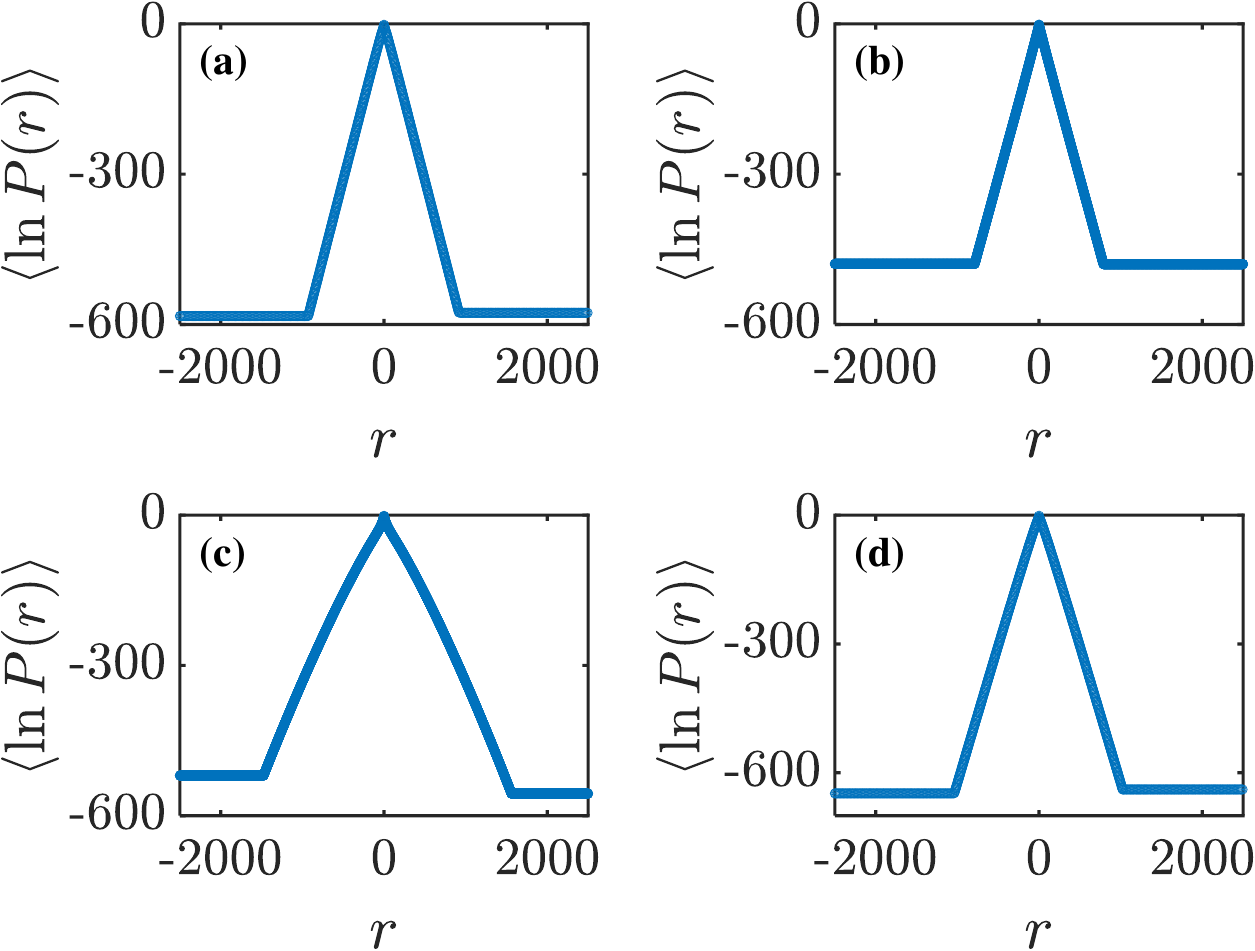}
	\caption{The disorder averaged the logarithm of the wave density $\langle \ln P(r)\rangle$ for (a) $t_1=0.29$ (TI), it exhibits exponentially decay. (b) $t_1=1.12$ (TAI), it is also exponentially decay. (c) $t_1=1.40$ (CP), the wave packet exhibits non-stationary behavior. (d) $t_1=1.90$ (NI), it shows exponential decay.  Other parameters are $t_2=1$, $N=5000$, $N_d=10^4$, $W=1.43$. The evolution time is $T=10^4$. Here we consider the periodic boundary condition. (Data in the plateaus are absent because the density $\langle\ln P(r)\rangle$ drops below the floating-point precision floor).}
	\label{N=5000}
\end{figure}
\begin{figure}[tbp]
	\centering
\includegraphics[width=0.48\textwidth]{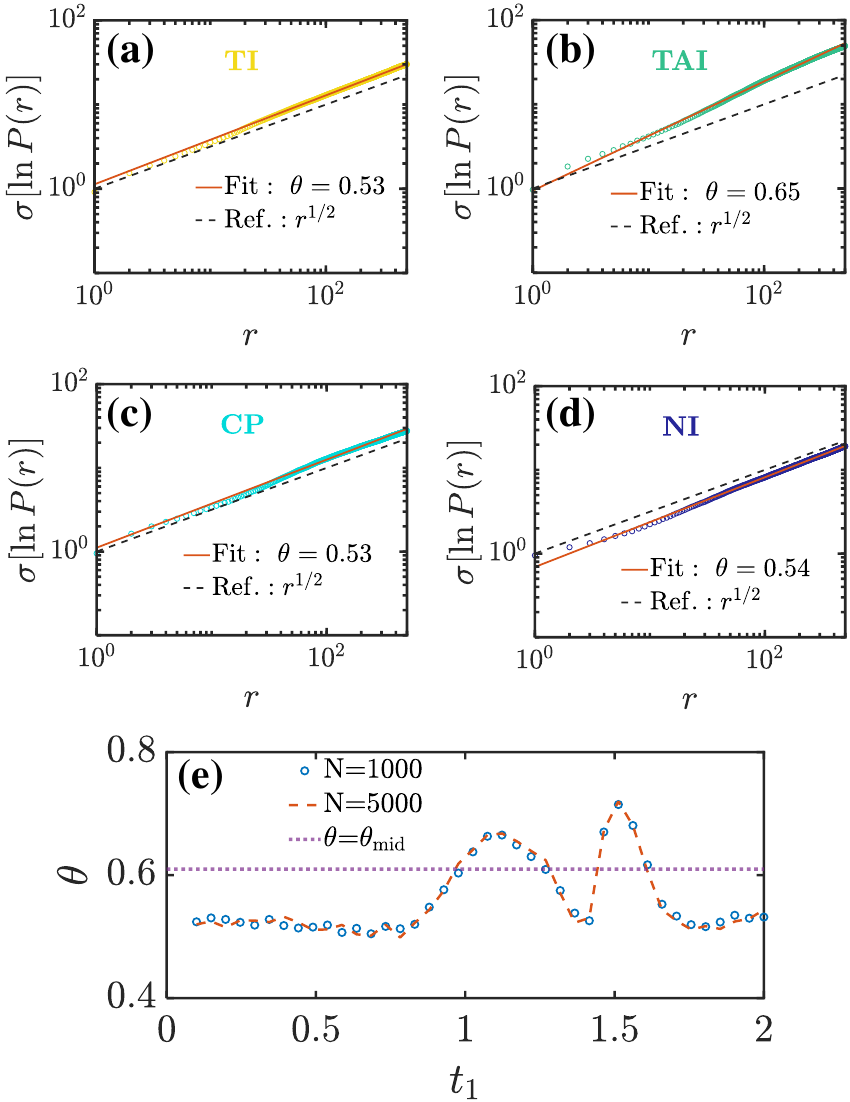}
	\caption{The fluctuation of the logarithm of the wave density $\ln P(r)$ with distance $r$ for $N=5000$. The blue circles represent the numerical data, the solid red line shows a power-law fit, the black dashed line indicates the  scaling $y = r^{1/2}$ for reference.  (a) $t_1=0.29$ (TI), the growth exponent is 0.53. (b) $t_1=1.12$ (TAI), the growth exponent is 0.65. (c) $t_1=1.40$ (CP) at the transition point, the growth exponent is 0.53. (d) $t_1=1.90$ (NI), the growth exponent is 0.54. (e) Fitted exponent $\theta$ as a function of $t_1$ for $N=1000$ and $N=5000$. The two curves are nearly identical over the entire parameter range. The horizontal dotted lines denote the midpoint thresholds $\theta_{\mathrm{mid}}(N)=0.61$. The portions above the corresponding thresholds operationally define the two fluctuation-enhanced regions, approximately $t_1\in[0.98,1.27]$ and $[1.44,1.61]$ for both system sizes. Other parameters are, $t_2=1$,  $N_d=10^4$, $W=1.43$. The evolution time is $T=10^4$. Here we consider the periodic boundary condition.}
	\label{N=5000_2}
\end{figure}
To test the sensitivity of the extracted exponent to the system size, we set the number of unit cells to $N=5000$. The logarithm of the wave density $\ln P(r)$ for different $t_1$ is shown in Fig.~\ref{N=5000} and Fig.~\ref{N=5000_2}, the results agree with those obtained for $N=1000$. 

For the topological and the trivial insulator phase away from the phase transition boundary, the wave packets exponentially decay for long evolution time. 
The growth exponents for the fluctuation of $\ln P(r)$ are close to $1/2$. For $t_1=1.12$ near to the transition, although $P(r)$ still exponentially decay, the growth exponent deviates the Gauss class and increases to $0.65$. For the transition point in Fig.~\ref{N=5000}(c), the wave packet is not exponentially decay. The fluctuation of logarithm is shown in Fig.~\ref{N=5000_2}(c), the growth exponent is $0.53$. The results for $N=5000$ are consistent with those for $N=1000$, indicating that our findings are free from finite-size effects. 

We further calculated $\theta(t_1)$ for $N=5000$ and compared it with the original result for $N=1000$, as shown in Fig.~\ref{N=5000_2}(e). Because $\theta$ varies continuously with $t_1$, the fluctuation-enhanced region does not have a unique sharp boundary. Moreover, the numerical values away from the enhancement are not exactly $1/2$ because of finite-size, finite-time, and fitting-window effects. We therefore introduce an operational midpoint threshold for each system size,
\begin{equation}
\theta_{\mathrm{mid}}(N)
=
\frac{\theta_{\min}(N)+\theta_{\max}(N)}{2},
\tag{C1}
\end{equation}
where $\theta_{\min}(N)$ and $\theta_{\max}(N)$ are, respectively, the minimum and maximum values of $\theta(t_1,N)$ over the parameter range shown in Fig.~\ref{N=5000_2}(e). The intervals satisfying
$\theta(t_1,N)\geq\theta_{\mathrm{mid}}(N)$ are operationally identified as the fluctuation-enhanced regions. Since $\theta$ decreases again at the transition itself, this criterion yields two separate intervals on the two sides of the transition. 

For $N=1000$, we obtain $\theta_{\min}=0.5046$ and $\theta_{\max}=0.7147$, giving $\theta_{\mathrm{mid}}=0.6097$. For $N=5000$, the corresponding values are $\theta_{\min}=0.4985$, $\theta_{\max}=0.7199$, and $\theta_{\mathrm{mid}}=0.6092$. Thus, $\theta_{\mathrm{mid}}\approx0.61$. 
The two thresholds are therefore nearly identical and visually overlap in Fig.~\ref{N=5000_2}(e). Applying this criterion, we obtain two fluctuation-enhanced intervals, approximately $t_1\in[0.98,1.27]$ and $t_1\in[1.44,1.61]$, with apparent widths of approximately $0.29$ and $0.17$, respectively. The intervals extracted for $N=1000$ and $N=5000$ are nearly identical. Thus, over the system sizes considered, no appreciable narrowing or broadening of the fluctuation-enhanced regions is observed.

\newpage
\bibliography{ref}

@PREAMBLE{
 "\providecommand{\noopsort}[1]{}" 
 # "\providecommand{\singleletter}[1]{#1}%" 
}

@article{anderson1958absence,
  title={Absence of diffusion in certain random lattices},
  author={Anderson, Philip W},
  journal={Physical review},
  volume={109},
  number={5},
  pages={1492},
  year={1958},
  publisher={APS}
}

@article{lagendijk2009fifty,
  title={Fifty years of Anderson localization},
  author={Lagendijk, Ad and Tiggelen, Bart van and Wiersma, Diederik S},
  journal={Physics today},
  volume={62},
  number={8},
  pages={24--29},
  year={2009},
  publisher={AIP Publishing}
}

@article{abrahams1979scaling,
  title={Scaling theory of localization: Absence of quantum diffusion in two dimensions},
  author={Abrahams, Elihu and Anderson, Philip W and Licciardello, Donald C and Ramakrishnan, Tiruppattur V},
  journal={Physical Review Letters},
  volume={42},
  number={10},
  pages={673},
  year={1979},
  publisher={APS}
}

@article{kramer1993localization,
  title={Localization: theory and experiment},
  author={Kramer, Bernhard and MacKinnon, Angus},
  journal={Reports on Progress in Physics},
  volume={56},
  number={12},
  pages={1469},
  year={1993},
  publisher={IOP Publishing}
}

@article{evers2008anderson,
  title={Anderson transitions},
  author={Evers, Ferdinand and Mirlin, Alexander D},
  journal={Reviews of Modern Physics},
  volume={80},
  number={4},
  pages={1355--1417},
  year={2008},
  publisher={APS}
}

@article{soukoulis1981off,
  title={Off-diagonal disorder in one-dimensional systems},
  author={Soukoulis, CM and Economou, EN},
  journal={Physical Review B},
  volume={24},
  number={10},
  pages={5698},
  year={1981},
  publisher={APS}
}

@article{mondragon2014topological,
  title = {Topological Criticality in the Chiral-Symmetric AIII Class at Strong Disorder},
  author = {Mondragon-Shem, Ian and Hughes, Taylor L. and Song, Juntao and Prodan, Emil},
  journal = {Phys. Rev. Lett.},
  volume = {113},
  issue = {4},
  pages = {046802},
  numpages = {5},
  year = {2014},
  month = {Jul},
  publisher = {American Physical Society},
  doi = {10.1103/PhysRevLett.113.046802},
  url = {https://link.aps.org/doi/10.1103/PhysRevLett.113.046802}
}

@article{de2016generalized,
  title={Generalized Dyson model: Nature of the zero mode and its implication in dynamics},
  author={De Tomasi, Giuseppe and Roy, Sthitadhi and Bera, Soumya},
  journal={Physical Review B},
  volume={94},
  number={14},
  pages={144202},
  year={2016},
  publisher={APS}
}

@article{dyson1953dynamics,
  title={The dynamics of a disordered linear chain},
  author={Dyson, Freeman J},
  journal={Physical Review},
  volume={92},
  number={6},
  pages={1331},
  year={1953},
  publisher={APS}
}

@article{weissmann1975density,
  title={Density of states of a one-dimensional system with off-diagonal disorder},
  author={Weissmann, M and Cohan, NV},
  journal={Journal of Physics C: Solid State Physics},
  volume={8},
  number={9},
  pages={L145},
  year={1975},
  publisher={IOP Publishing}
}

@article{bush1975anomalous,
  title={On the anomalous behaviour at the band centre of a one-dimensional system with off-diagonal disorder},
  author={Bush, RL},
  journal={Journal of Physics C: Solid State Physics},
  volume={8},
  number={23},
  pages={L547},
  year={1975},
  publisher={IOP Publishing}
}

@article{theodorou1976extended,
  title={Extended states in a one-demensional system with off-diagonal disorder},
  author={Theodorou, George and Cohen, Morrel H},
  journal={Physical Review B},
  volume={13},
  number={10},
  pages={4597},
  year={1976},
  publisher={APS}
}

@article{eggarter1978singular,
  title={Singular behavior of tight-binding chains with off-diagonal disorder},
  author={Eggarter, TP and Riedinger, R},
  journal={Physical Review B},
  volume={18},
  number={2},
  pages={569},
  year={1978},
  publisher={APS}
}

@article{mirlin2000statistics,
  title={Statistics of energy levels and eigenfunctions in disordered systems},
  author={Mirlin, Alexander D},
  journal={Physics Reports},
  volume={326},
  number={5-6},
  pages={259--382},
  year={2000},
  publisher={Elsevier}
}

@article{anderson1980new,
  title={New method for a scaling theory of localization},
  author={Anderson, Philip W and Thouless, DJ and Abrahams, E and Fisher, DS},
  journal={Physical Review B},
  volume={22},
  number={8},
  pages={3519},
  year={1980},
  publisher={APS}
}

@article{furstenberg1963noncommuting,
  title={Noncommuting random products},
  author={Furstenberg, Harry},
  journal={Transactions of the American Mathematical Society},
  volume={108},
  number={3},
  pages={377--428},
  year={1963},
  publisher={JSTOR}
}

@article{li2009topological,
  title={Topological anderson insulator},
  author={Li, Jian and Chu, Rui-Lin and Jain, Jainendra K and Shen, Shun-Qing},
  journal={Physical review letters},
  volume={102},
  number={13},
  pages={136806},
  year={2009},
  publisher={APS}
}

@article{groth2009theory,
  title={Theory of the topological Anderson insulator},
  author={Groth, CW and Wimmer, M and Akhmerov, AR and Tworzyd{\l}o, J and Beenakker, CWJ},
  journal={Physical review letters},
  volume={103},
  number={19},
  pages={196805},
  year={2009},
  publisher={APS}
}

@article{jiang2009numerical,
  title={Numerical study of the topological Anderson insulator in HgTe/CdTe quantum wells},
  author={Jiang, Hua and Wang, Lei and Sun, Qing-feng and Xie, XC},
  journal={Physical Review B—Condensed Matter and Materials Physics},
  volume={80},
  number={16},
  pages={165316},
  year={2009},
  publisher={APS}
}

@article{guo2010topological,
  title={Topological Anderson insulator in three dimensions},
  author={Guo, H-M and Rosenberg, G and Refael, G and Franz, M},
  journal={Physical review letters},
  volume={105},
  number={21},
  pages={216601},
  year={2010},
  publisher={APS}
}

@article{altland2014quantum,
  title={Quantum criticality of quasi-one-dimensional topological Anderson insulators},
  author={Altland, Alexander and Bagrets, Dmitry and Fritz, Lars and Kamenev, Alex and Schmiedt, Hanno},
  journal={Physical Review Letters},
  volume={112},
  number={20},
  pages={206602},
  year={2014},
  publisher={APS}
}

@article{su1979solitons,
  title={Solitons in polyacetylene},
  author={Su, Wu-Pei and Schrieffer, John Robert and Heeger, Alan J},
  journal={Physical review letters},
  volume={42},
  number={25},
  pages={1698},
  year={1979},
  publisher={APS}
}

@article{meier2018observation,
  title={Observation of the topological Anderson insulator in disordered atomic wires},
  author={Meier, Eric J and An, Fangzhao Alex and Dauphin, Alexandre and Maffei, Maria and Massignan, Pietro and Hughes, Taylor L and Gadway, Bryce},
  journal={Science},
  volume={362},
  number={6417},
  pages={929--933},
  year={2018},
  publisher={American Association for the Advancement of Science}
}

@article{stutzer2018photonic,
  title={Photonic topological Anderson insulators},
  author={St{\"u}tzer, Simon and Plotnik, Yonatan and Lumer, Yaakov and Titum, Paraj and Lindner, Netanel H and Segev, Mordechai and Rechtsman, Mikael C and Szameit, Alexander},
  journal={Nature},
  volume={560},
  number={7719},
  pages={461--465},
  year={2018},
  publisher={Nature Publishing Group UK London}
}

@article{zhang2021experimental,
  title={Experimental observation of higher-order topological Anderson insulators},
  author={Zhang, Weixuan and Zou, Deyuan and Pei, Qingsong and He, Wenjing and Bao, Jiacheng and Sun, Houjun and Zhang, Xiangdong},
  journal={Physical Review Letters},
  volume={126},
  number={14},
  pages={146802},
  year={2021},
  publisher={APS}
}

@article{liu2020topological,
  title={Topological Anderson insulator in disordered photonic crystals},
  author={Liu, Gui-Geng and Yang, Yihao and Ren, Xin and Xue, Haoran and Lin, Xiao and Hu, Yuan-Hang and Sun, Hong-xiang and Peng, Bo and Zhou, Peiheng and Chong, Yidong and others},
  journal={Physical Review Letters},
  volume={125},
  number={13},
  pages={133603},
  year={2020},
  publisher={APS}
}

@article{lin2022observation,
  title={Observation of non-Hermitian topological Anderson insulator in quantum dynamics},
  author={Lin, Quan and Li, Tianyu and Xiao, Lei and Wang, Kunkun and Yi, Wei and Xue, Peng},
  journal={Nature Communications},
  volume={13},
  number={1},
  pages={3229},
  year={2022},
  publisher={Nature Publishing Group UK London}
}

@article{chen2024realization,
  title={Realization of Time-Reversal invariant photonic topological Anderson insulators},
  author={Chen, Xiao-Dong and Gao, Zi-Xuan and Cui, Xiaohan and Mo, Hao-Chang and Chen, Wen-Jie and Zhang, Ruo-Yang and Chan, CT and Dong, Jian-Wen},
  journal={Physical Review Letters},
  volume={133},
  number={13},
  pages={133802},
  year={2024},
  publisher={APS}
}

@article{ren2024realization,
  title={Realization of gapped and ungapped photonic topological Anderson insulators},
  author={Ren, Mina and Yu, Ye and Wu, Bintao and Qi, Xin and Wang, Yiwei and Yao, Xiaogang and Ren, Jie and Guo, Zhiwei and Jiang, Haitao and Chen, Hong and others},
  journal={Physical Review Letters},
  volume={132},
  number={6},
  pages={066602},
  year={2024},
  publisher={APS}
}

@article{mott1961theory,
  title={The theory of impurity conduction},
  author={Mott, Nevill F and Twose, WD},
  journal={Advances in physics},
  volume={10},
  number={38},
  pages={107--163},
  year={1961},
  publisher={Taylor \& Francis}
}

@article{borland1963nature,
  title={The nature of the electronic states in disordered one-dimensional systems},
  author={Borland, RE},
  journal={Proceedings of the Royal Society of London. Series A. Mathematical and Physical Sciences},
  volume={274},
  number={1359},
  pages={529--545},
  year={1963},
  publisher={The Royal Society London}
}

@book{asboth2016short,
  title={A short course on topological insulators},
  author={Asb{\'o}th, J{\'a}nos K and Oroszl{\'a}ny, L{\'a}szl{\'o} and P{\'a}lyi, Andr{\'a}s},
  volume={919},
  year={2016},
  publisher={Springer}
}

@article{roati2008anderson,
  title={Anderson localization of a non-interacting Bose--Einstein condensate},
  author={Roati, Giacomo and D’Errico, Chiara and Fallani, Leonardo and Fattori, Marco and Fort, Chiara and Zaccanti, Matteo and Modugno, Giovanni and Modugno, Michele and Inguscio, Massimo},
  journal={Nature},
  volume={453},
  number={7197},
  pages={895--898},
  year={2008},
  publisher={Nature Publishing Group UK London}
}

@article{billy2008direct,
  title={Direct observation of Anderson localization of matter waves in a controlled disorder},
  author={Billy, Juliette and Josse, Vincent and Zuo, Zhanchun and Bernard, Alain and Hambrecht, Ben and Lugan, Pierre and Cl{\'e}ment, David and Sanchez-Palencia, Laurent and Bouyer, Philippe and Aspect, Alain},
  journal={Nature},
  volume={453},
  number={7197},
  pages={891--894},
  year={2008},
  publisher={Nature Publishing Group}
}

@article{mu2024kardar,
  title = {Kardar-Parisi-Zhang Physics in the Density Fluctuations of Localized Two-Dimensional Wave Packets},
  author = {Mu, Sen and Gong, Jiangbin and Lemari\'e, Gabriel},
  journal = {Phys. Rev. Lett.},
  volume = {132},
  issue = {4},
  pages = {046301},
  numpages = {6},
  year = {2024},
  month = {Jan},
  publisher = {American Physical Society},
  doi = {10.1103/PhysRevLett.132.046301},
  url = {https://link.aps.org/doi/10.1103/PhysRevLett.132.046301}
}

@article{izem2025kardar,
      title={Kardar-Parisi-Zhang and glassy properties in 2D Anderson localization: eigenstates and wave packets}, 
      author={Noam Izem and Bertrand Georgeot and Jiangbin Gong and Gabriel Lemarié and Sen Mu},
      year={2025},
      journal={arXiv:2512.12085},
      primaryClass={cond-mat.dis-nn},
      url={https://arxiv.org/abs/2512.12085}, 
}

@article{tang2020topological,
  title={Topological Anderson insulators in two-dimensional non-Hermitian disordered systems},
  author={Tang, Ling-Zhi and Zhang, Ling-Feng and Zhang, Guo-Qing and Zhang, Dan-Wei},
  journal={Physical Review A},
  volume={101},
  number={6},
  pages={063612},
  year={2020},
  publisher={APS}
}

@article{ziman1982localization,
  title={Localization with off-diagonal disorder: A qualitative theory},
  author={Ziman, Timothy AL},
  journal={Physical Review B},
  volume={26},
  number={12},
  pages={7066},
  year={1982},
  publisher={APS}
}

@article{hilke2025disordered,
  title={The disordered Su-Schrieffer-Heeger model},
  author={Hilke, Michael},
  journal={arXiv preprint arXiv:2512.24738},
  year={2025}
}

@article{tang2022topological,
  title={Topological Anderson insulators with different bulk states in quasiperiodic chains},
  author={Tang, Ling-Zhi and Liu, Shu-Na and Zhang, Guo-Qing and Zhang, Dan-Wei},
  journal={Physical Review A},
  volume={105},
  number={6},
  pages={063327},
  year={2022},
  publisher={APS}
}

@article{somoza2007universal,
  title={Universal distribution functions in two-dimensional localized systems},
  author={Somoza, AM and Ortuno, M and Prior, J},
  journal={Physical Review Letters},
  volume={99},
  number={11},
  pages={116602},
  year={2007},
  publisher={APS}
}

@article{lemarie2019glassy,
  title={Glassy properties of Anderson localization: Pinning, avalanches, and chaos},
  author={Lemari{\'e}, Gabriel},
  journal={Physical review letters},
  volume={122},
  number={3},
  pages={030401},
  year={2019},
  publisher={APS}
}

@article{kardar1986dynamic,
  title={Dynamic scaling of growing interfaces},
  author={Kardar, Mehran and Parisi, Giorgio and Zhang, Yi-Cheng},
  journal={Physical Review Letters},
  volume={56},
  number={9},
  pages={889},
  year={1986},
  publisher={APS}
}

@article{prior2005conductance,
  title={Conductance fluctuations and single-parameter scaling in two-dimensional disordered systems},
  author={Prior, J and Somoza, AM and Ortuno, M},
  journal={Physical Review B—Condensed Matter and Materials Physics},
  volume={72},
  number={2},
  pages={024206},
  year={2005},
  publisher={APS}
}

@article{prior2009conductance,
  title={Conductance distribution in two-dimensional localized systems with and without magnetic fields},
  author={Prior, J and Somoza, AM and Ortuno, M},
  journal={The European Physical Journal B},
  volume={70},
  number={4},
  pages={513--521},
  year={2009},
  publisher={Springer}
}

@article{somoza2015unbinding,
  title={Unbinding transition in semi-infinite two-dimensional localized systems},
  author={Somoza, AM and Le Doussal, P and Ortuno, M},
  journal={Physical Review B},
  volume={91},
  number={15},
  pages={155413},
  year={2015},
  publisher={APS}
}

@article{brouwer2000density,
  title={Density of states in coupled chains with off-diagonal disorder},
  author={Brouwer, Piet W and Mudry, Christopher and Furusaki, Akira},
  journal={Physical Review Letters},
  volume={84},
  number={13},
  pages={2913},
  year={2000},
  publisher={APS}
}

@article{balents1997delocalization,
  title={Delocalization transition via supersymmetry in one dimension},
  author={Balents, Leon and Fisher, Matthew PA},
  journal={Physical Review B},
  volume={56},
  number={20},
  pages={12970},
  year={1997},
  publisher={APS}
}

@article{beenakker1997random,
  title={Random-matrix theory of quantum transport},
  author={Beenakker, Carlo WJ},
  journal={Reviews of modern physics},
  volume={69},
  number={3},
  pages={731},
  year={1997},
  publisher={APS}
}

@article{dorokhov1982transmission,
  title={Transmission coefficient and the localization length of an electron in N bound disordered chains},
  author={Dorokhov, ON},
  journal={Soviet Journal of Experimental and Theoretical Physics Letters},
  volume={36},
  pages={318},
  year={1982}
}

@article{mello1988macroscopic,
  title={Macroscopic approach to multichannel disordered conductors},
  author={Mello, PA and Pereyra, P and Kumar, N},
  journal={Annals of Physics},
  volume={181},
  number={2},
  pages={290--317},
  year={1988},
  publisher={Elsevier}
}

@book{mello2004quantum,
  title={Quantum Transport in Mesoscopic Systems: Complexity and Statistical Fluctuations. A Maximum Entropy Viewpoint},
  author={Mello, Pier A and Kumar, Narendra},
  year={2004},
  publisher={Oxford University Press}
}

@book{mehta2004random,
  title={Random matrices},
  author={Mehta, Madan Lal},
  volume={142},
  year={2004},
  publisher={Elsevier}
}

@article{muttalib1999generalized,
  title={Generalized Fokker-Planck equation for multichannel disordered quantum conductors},
  author={Muttalib, KA and Klauder, JR},
  journal={Physical review letters},
  volume={82},
  number={21},
  pages={4272},
  year={1999},
  publisher={APS}
}

@article{muttalib2002generalization,
  title={Generalization of the DMPK equation beyond quasi one dimension},
  author={Muttalib, KA and Gopar, V{\'\i}ctor A},
  journal={Physical Review B},
  volume={66},
  number={11},
  pages={115318},
  year={2002},
  publisher={APS}
}

@article{douglas2014generalized,
  title={The generalized DMPK equation revisited: towards a systematic derivation},
  author={Douglas, Andrew and Marko{\v{s}}, Peter and Muttalib, KA},
  journal={Journal of Physics A: Mathematical and Theoretical},
  volume={47},
  number={12},
  pages={125103},
  year={2014},
  publisher={IOP Publishing}
}

@article{suslov2018general,
  title={General form of DMPK equation},
  author={Suslov, IM},
  journal={Journal of Experimental and Theoretical Physics},
  volume={127},
  number={1},
  pages={131--142},
  year={2018},
  publisher={Springer}
}

@article{li2024mapping,
  title={Mapping the topology-localization phase diagram with quasiperiodic disorder using a programmable superconducting simulator},
  author={Li, Xuegang and Xu, Huikai and Wang, Junhua and Tang, Ling-Zhi and Zhang, Dan-Wei and Yang, Chuhong and Su, Tang and Wang, Chenlu and Mi, Zhenyu and Sun, Weijie and others},
  journal={Physical Review Research},
  volume={6},
  number={4},
  pages={L042038},
  year={2024},
  publisher={APS}
}

@article{khudaiberdiev2025two,
  title={Two-dimensional topological Anderson insulator in a HgTe-based semimetal},
  author={Khudaiberdiev, DA and Kvon, ZD and Ryzhkov, MS and Kozlov, DA and Mikhailov, NN and Pimenov, A},
  journal={Physical Review Research},
  volume={7},
  number={2},
  pages={L022033},
  year={2025},
  publisher={APS}
}

@article{li2025demonstration,
  title={Demonstration of Time-reversal Symmetric Two-Dimensional Photonic Topological Anderson Insulator},
  author={Li, Zhe and Chen, Ziming and Kong, Deyang and Li, Yongzhuo and Cui, Kaiyu and Feng, Xue and Huang, Yidong},
  journal={arXiv preprint arXiv:2501.11251},
  year={2025}
}

@article{mannai2026realization,
  title={Realization of staircase topological Anderson phase transitions},
  author={Mannai, Marwa and Shu, Yaoyao and Haddad, Sonia and Ren, Mina and Chen, Hong and Sun, Yong and Sati, Hisham},
  journal={arXiv preprint arXiv:2601.14769},
  year={2026}
}

@article{zangeneh2020disorder,
  title={Disorder-induced signal filtering with topological metamaterials},
  author={Zangeneh-Nejad, Farzad and Fleury, Romain},
  journal={Advanced Materials},
  volume={32},
  number={28},
  pages={2001034},
  year={2020},
  publisher={Wiley Online Library}
}

@article{liu2021acoustic,
  title={Acoustic topological Anderson insulators},
  author={Liu, Hui and Xie, Boyang and Wang, Haonan and Liu, Wenwei and Li, Zhancheng and Cheng, Hua and Tian, Jianguo and Liu, Zhengyou and Chen, Shuqi},
  journal={arXiv preprint arXiv:2111.06520},
  year={2021}
}

@article{pbrz-3lrl,
  title = {Two-dimensional Anderson localization and KPZ subuniversality classes: Sensitivity to boundary conditions and insensitivity to symmetry classes},
  author = {Swain, Nyayabanta and Adam, Shaffique and Lemari\'e, Gabriel},
  journal = {Phys. Rev. Res.},
  volume = {8},
  issue = {1},
  pages = {013038},
  numpages = {18},
  year = {2026},
  month = {Jan},
  publisher = {American Physical Society},
  doi = {10.1103/pbrz-3lrl},
  url = {https://link.aps.org/doi/10.1103/pbrz-3lrl}
}

@article{7d45-yrwq,
  title = {Universal fluctuations of two localized interacting particles in one dimension},
  author = {Mu, Sen and Lemari\'e, Gabriel and Gong, Jiangbin},
  journal = {Phys. Rev. B},
  volume = {112},
  issue = {1},
  pages = {014201},
  numpages = {12},
  year = {2025},
  month = {Jul},
  publisher = {American Physical Society},
  doi = {10.1103/7d45-yrwq},
  url = {https://link.aps.org/doi/10.1103/7d45-yrwq}
}

@article{krishna2020beyond,
  title={Beyond universal behavior in the one-dimensional chain with random nearest-neighbor hopping},
  author={Krishna, Akshay and Bhatt, Ravindra N},
  journal={Physical Review B},
  volume={101},
  number={22},
  pages={224203},
  year={2020},
  publisher={APS}
}

@article{fleishman1977fluctuations,
  title={Fluctuations and localization in one dimension},
  author={Fleishman, L and Licciardello, DC},
  journal={Journal of Physics C: Solid State Physics},
  volume={10},
  number={6},
  pages={L125--L126},
  year={1977}
}

@article{motrunich2002particle,
  title={Particle-hole symmetric localization in two dimensions},
  author={Motrunich, Olexei and Damle, Kedar and Huse, David A},
  journal={Physical Review B},
  volume={65},
  number={6},
  pages={064206},
  year={2002},
  publisher={APS}
}

@article{gade1993anderson,
  title={Anderson localization for sublattice models},
  author={Gade, Renate},
  journal={Nuclear Physics B},
  volume={398},
  number={3},
  pages={499--515},
  year={1993},
  publisher={Elsevier}
}

@article{eilmes1998two,
  title={The two-dimensional Anderson model of localization with random hopping},
  author={Eilmes, Andrzej and R{\"o}mer, Rudolf A and Schreiber, Michael},
  journal={The European Physical Journal B-Condensed Matter and Complex Systems},
  volume={1},
  number={1},
  pages={29--38},
  year={1998},
  publisher={Springer}
}

@article{schweitzer2012scaling,
  title={Scaling at chiral quantum critical points in two dimensions},
  author={Schweitzer, L and Marko{\v{s}}, P},
  journal={Physical Review B—Condensed Matter and Materials Physics},
  volume={85},
  number={19},
  pages={195424},
  year={2012},
  publisher={APS}
}

@article{zuo2022reentrant,
  title={Reentrant localization transition in the Su-Schrieffer-Heeger model with random-dimer disorder},
  author={Zuo, Zheng-Wei and Kang, Dawei},
  journal={Physical Review A},
  volume={106},
  number={1},
  pages={013305},
  year={2022},
  publisher={APS}
}

@article{nava2023lindblad,
  title={Lindblad master equation approach to the topological phase transition in the disordered Su-Schrieffer-Heeger model},
  author={Nava, Andrea and Campagnano, Gabriele and Sodano, Pasquale and Giuliano, Domenico},
  journal={Physical Review B},
  volume={107},
  number={3},
  pages={035113},
  year={2023},
  publisher={APS}
}

@article{cinnirella2024fate,
  title={Fate of high winding number topological phases in the disordered extended Su-Schrieffer-Heeger model},
  author={Cinnirella, Emmanuele G and Nava, Andrea and Campagnano, Gabriele and Giuliano, Domenico},
  journal={Physical Review B},
  volume={109},
  number={3},
  pages={035114},
  year={2024},
  publisher={APS}
}

@article{wang2025emergent,
  title={Emergent topological re-entrant phase transition in a generalized quasiperiodic modulated Su-Schrieffer-Heeger model},
  author={Wang, Xiao-Ming and Li, Shan-Zhong and Li, Zhi},
  journal={Physical Review A},
  volume={111},
  number={2},
  pages={022214},
  year={2025},
  publisher={APS}
}

@article{sircar2025disorder,
  title={Disorder driven topological phase transitions in 1D mechanical quasicrystals},
  author={Sircar, Sayan},
  journal={Physics Letters A},
  pages={131192},
  year={2025},
  publisher={Elsevier}
}

@article{zhang2026reentrant,
  title={Reentrant topological phase transition in a periodically modulated Su-Schrieffer-Heeger model},
  author={Zhang, Xin-Yue and Nie, Xiao-Feng and Zhang, Jie and Zhong, Li-Nan and Cao, Ji and Wang, Hong-Fu},
  journal={Physical Review A},
  volume={113},
  number={4},
  pages={042219},
  year={2026},
  publisher={APS}
}

@article{roy2021reentrant,
  title={Reentrant localization transition in a quasiperiodic chain},
  author={Roy, Shilpi and Mishra, Tapan and Tanatar, Bilal and Basu, Saurabh},
  journal={Physical Review Letters},
  volume={126},
  number={10},
  pages={106803},
  year={2021},
  publisher={APS}
}

@article{wu2016disorder,
  title={Disorder effects in topological states: Brief review of the recent developments},
  author={Wu, Binglan and Song, Juntao and Zhou, Jiaojiao and Jiang, Hua},
  journal={Chinese Physics B},
  volume={25},
  number={11},
  pages={117311},
  year={2016},
  publisher={IOP Publishing}
}

\end{document}